\documentclass{bmcart}
\usepackage{amsfonts}
\usepackage{amssymb}
\usepackage[all]{xy}
\usepackage{pdfpages}
\usepackage{tikz}
\usepackage{mathtools,xparse}
\usepackage{amsthm,amsmath}

\usepackage[utf8]{inputenc} 

\DeclarePairedDelimiterX{\bra}[1]{\langle}{\rvert}{#1\,}
\DeclarePairedDelimiterX{\ket}[1]{\lvert}{\rangle}{\,#1}
\DeclarePairedDelimiterX{\makebraket}[1]{\langle}{\rangle}{#1}

\NewDocumentCommand{\braket}{som}{%
  \begingroup\activatebraketbar
  \IfBooleanTF{#1}
    {\makebraket*{#3}}
    {\IfNoValueTF{#2}{\makebraket{#3}}{\makebraket[#2]{#3}}}%
  \endgroup
}

\makeatletter
\newcommand{\braketbar}{%
  \,\delimsize\vert\@ifnextchar|{\!}{\,}%
}
\makeatother
\newcommand{\activatebraketbar}{%
  \begingroup\lccode`~=`|\lowercase{\endgroup\let~}\braketbar
  \mathcode`|="8000
}

\numberwithin{equation}{section}
\newtheorem{theorem}{Theorem}[section]

\newtheorem{definition}[theorem]{Definition}

\numberwithin{equation}{section}

\usepgflibrary{shapes}
\usetikzlibrary{backgrounds}
\usetikzlibrary{arrows}
\usetikzlibrary{shapes,shapes.geometric,shapes.misc}

\tikzstyle{tikzfig}=[baseline=-0.25em,scale=0.5]

\pgfkeys{/tikz/tikzit fill/.initial=0}
\pgfkeys{/tikz/tikzit draw/.initial=0}
\pgfkeys{/tikz/tikzit shape/.initial=0}
\pgfkeys{/tikz/tikzit category/.initial=0}

\pgfdeclarelayer{edgelayer}
\pgfdeclarelayer{nodelayer}
\pgfsetlayers{background,edgelayer,nodelayer,main}

\tikzstyle{none}=[inner sep=0mm]

\newcommand{\tikzfig}[1]{%
{\tikzstyle{every picture}=[tikzfig]
\IfFileExists{#1.tikz}
  {\input{#1.tikz}}
  {%
    \IfFileExists{./figures/#1.tikz}
      {\input{./figures/#1.tikz}}
      {\tikz[baseline=-0.5em]{\node[draw=red,font=\color{red},fill=red!10!white] {\textit{#1}};}}%
  }}%
}

\tikzstyle{every loop}=[]

\tikzstyle{new style 0}=[fill=black, draw=black, shape=circle, tikzit fill=black, tikzit draw=black]
\tikzstyle{new style 1}=[fill=white, draw=black, shape=circle, tikzit fill=white, tikzit draw=black]
\tikzstyle{black dot}=[fill={rgb,255: red,128; green,128; blue,128}, draw=black, shape=circle, tikzit fill={rgb,255: red,128; green,128; blue,128}, tikzit draw=black]
\tikzstyle{white dot}=[fill=white, draw=black, shape=circle, tikzit fill=white, tikzit draw=black]
\tikzstyle{new style 2}=[fill={rgb,255: red,128; green,128; blue,128}, draw=black, shape=circle, tikzit fill={rgb,255: red,128; green,128; blue,128}, tikzit draw=black]
\tikzstyle{new style 3}=[fill=white, draw=black, shape=circle, tikzit fill=white, tikzit draw=black]

\startlocaldefs
\endlocaldefs

\begin{document}

\begin{frontmatter}

\begin{fmbox}
\dochead{Research}


\title{Parametrised Quantum Circuits of Synonymous Sentences in Quantum Natural Language Processing}


\author[
  addressref={aff1},                   
  email={mina.abbaszade@math.uk.ac.ir}   
]{\inits{M.}\fnm{Mina} \snm{Abbaszadeh}}
\author[
  addressref={aff2},
]{\inits{S.}\fnm{Seyyed Shahin} \snm{Mousavi}}
\author[
   addressref={aff1, aff3},
   corref={aff1},
  email={vahidsalari@iut.ac.ir}
]{\inits{V.}\fnm{Vahid} \snm{Salari}}


\address[id=aff1]{%
  \orgdiv{Department of Physics},
  \orgname{Isfahan University of Technology},
  \street{},
  \postcode{84156-83111}
  \city{Isfahan},
  \cny{Iran}
}

\address[id=aff2]{
  \orgdiv{Pure Mathematics Department},             
  \orgname{Shahid Bahonar University of Kerman},          
  \city{Kerman},                              
  \cny{Iran}                                    
}
\address[id=aff3]{%
  \orgdiv{Department of Physical Chemistry},
  \orgname{University of Basque Country UPV/EHU},
  \street{ Apdo. 644},
  \postcode{48080}
  \city{Bilbao},
  \cny{Spain}
}


\end{fmbox}


\begin{abstractbox}

\begin{abstract} 
	In this paper we develop a compositional vector-based semantics of positive transitive sentences in quantum natural language processing for a non-English language, i.e. Persian, to compare the parametrised quantum circuits of two synonymous sentences in two languages, English and Persian. By considering grammar+meaning of a transitive sentence, we translate DisCoCat diagram via ZX-calculus into quantum circuit form. 
	Also we use a bigraph method to rewrite DisCoCat diagram and turn into quantum circuit  in the semantic side. 
\end{abstract}




\end{abstractbox}
%

\end{frontmatter}

\section*{Introduction}
Natural language processing (NLP) is a subgroup of linguistics and artificial intelligence used for language interactions between computers and human, e.g. programming computers to analyze natural language data with large volumes. A computer can understand the meanings and concepts of the texts in documents, recognises speech, and generates natural language via NLP. In fact, NLP was proposed first in 1950 by Alan Turing \cite{Turing} i.e. now called the Turing test as a criterion of intelligence for automated interpretation and generation of natural language. Recently a group of researchers at OpenAI have developed Generative Pre-trained Transformer 3 (GPT-3) language model \cite{GPT3}, as the largest non-sparse language model with higher number of parameters and a higher level of accuracy versus previous models with capacity of ten times larger than that of Microsoft's Turing-NLG to date. On the other side, some quantum approaches for NLP have been developed that may reach some quantum advantages over classical counterparts in future \cite{circuit, MLST}.  Protocols for quantum Natural Language Processing (QNLP) have two aspects: semantic and syntax. Both aspects are performed by a mathematical framework.
Compact closed categories are used to provide semantics for quantum protocols  \cite{protocol}. 
The use of quantum maps for describing meaning in natural language was started by Bob Coecke \cite{math}.
Coecke has introduced diagrammatic language to speak about processes and how they compose \cite{p}.
The diagrammatic language of non-commutative categorical quantum logic represents reduction diagrams for sentences, and allows one to compare the grammatical structures of sentences in different languages. Sadrzadeh has used pregroups to provide an algebraic analysis of Persian sentences \cite{m}.	
Pregroups are used to encode the grammar of languages. One can fix a set of basic grammatical roles and a partial ordering between them, then freely can generate a pregroup of these types \cite{math}. The category of finite dimensional vector spaces and pregroups are monoidal categories. Models of the semantic of positive and negative transitive sentences are given in ref. \cite{math}. 
Moreover, Frobenius algebras are used to model the semantics of subject and object relative pronouns \cite{frob}.
Brian Tyrrell \cite{t} has used vector space distributional compositional categorical
models of meaning to compare the meaning of sentences in Irish and in English.
Here, we use vector-based models of semantic composition to model the semantics of positive transitive sentences in Persian. According to \cite{circuit} the DisCoCat diagram is simplified to some other diagram and is turned into a quantum circuit, which can be compiled via noisy intermediate-scale (NISQ) devices. The grammatical quantum circuits are spanned by a set $\theta$. The meaning of the words and hence whole sentence are encoded in the created semantic space. Finally, we rewrite the diagram as a bipartite graph to turn a quantum circuit.
ZX-calculus, like a translator, turns a linguistic diagram into a quantum circuit.  
According to \cite{f} we consider both grammar and meaning of a grammatical sentence in Persian and turn DisCoCat diagram into a quantum circuit form.

\section{Preliminaries}

	In this section, we provide some content, which will be used throughout this paper. See the references ~\cite{frob} and~\cite{math} for more details. 	
\begin{definition}
 A category $\mathcal{C}$ consists of:
\begin{itemize}
 \item 
 a class $obj(\mathcal{C})$, called the class of objects;
 \item
 for every two objects $A, B$ a class $\mathcal{C}(A,B)$ of morphisms; it is convenient to abbreviate $f \in \mathcal{C}(A,B)$ by $f : A \rightarrow B$;
   \item
 for every two morphisms $f \in \mathcal{C}(A,B)$ and $g \in \mathcal{C}(B,C)$, a morphism $g\circ f \in \mathcal{C}(A,C)$. These must satisfy the following properties, for all objects $A, B, C, D$
and all morphisms $f \in \mathcal{C}(A,B), g \in \mathcal{C}(B,C), h \in \mathcal{C}(C,D)$: 
$$h \circ (g \circ f) = (h \circ g) \circ f;$$
  \item
 for every object $A$ there is an identity morphism $1_A \in \mathcal{C}(A,A)$; for $f \in \mathcal{C}(A,B)$ we have 
 $$1_B \circ f = f = f \circ 1_A$$
 \end{itemize}
 \end{definition}
 
 \begin{definition}
 A monoidal category is a category $\mathcal{C}$ with the following properties:
 \begin{itemize}
  \item
  a functor $\otimes : \mathcal{C} \times \mathcal{C} \rightarrow \mathcal{C}$, called the tensor product and we have
  $$ (A \otimes B) \otimes C = A \otimes (B \otimes C);$$
  \item
  there is a unit object $I$ such that
  $$I \otimes A = A = A \otimes I;$$
  \item
  for each ordered pair morphisms $f \in \mathcal{C}(A,C), g \in \mathcal{C}(B,D)$ we have $f \otimes g : A \otimes B \rightarrow C \otimes D$ such that
  $$ (g_1 \otimes g_2) \circ (f_1 \otimes f_2)=(g_1 \circ f_1) \otimes (g_2 \circ f_2).$$
\end{itemize}
\end{definition}

Monoidal categories are used to encode semantic and syntax of sentences in different languages. 

\begin{definition}
A symmetric monoidal category is a monoidal category $\mathcal{C}$ such that the tensor product is symmetric. This means that there is a natural isomorphism $\eta$ such that for all objects $A, B \in \mathcal{C},$
$\xymatrix{A \otimes B \ar[r]^{\eta_{A,B}} & B \otimes A}$
is an isomorphism. 
\end{definition}

Graphical language is a high-level language for researching in quantum processes, which has applications in many areas such as QNLP and modelling quantum circuits.   

\subsection{Graphical language for monoidal category}	
According to \cite{math}, morphisms are depicted by boxes, with input and output wires. For example, the morphisms
$$1_A \ \ \ \ \  f \ \ \ \ \  g \circ f \ \ \ \ \  f \otimes g$$
where $ f : A \rightarrow B$ and  $g : B \rightarrow C, $ are depicted as follows:
\vspace{1cm}
\begin{center}
\begin{tikzpicture}
	\begin{pgfonlayer}{nodelayer}
		\node [style=none] (0) at (-4.75, 1.5) {};
		\node [style=none] (1) at (-4.75, -0.25) {};
		\node [style=none] (2) at (-4.5, -0.25) {};
		\node [style=none] (3) at (-4.5, -0.25) {A};
		\node [style=none] (4) at (-3, 1.75) {};
		\node [style=none] (5) at (-3, 1) {};
		\node [style=none] (6) at (-2, 1.75) {};
		\node [style=none] (7) at (-2, 1) {};
		\node [style=none] (8) at (-2.5, 1.75) {};
		\node [style=none] (9) at (-2.5, 1) {};
		\node [style=none] (10) at (-2.5, -0.25) {};
		\node [style=none] (11) at (-2.5, 3) {};
		\node [style=none] (12) at (-2.5, 1.5) {};
		\node [style=none] (13) at (-2.5, 1.25) {$f$};
		\node [style=none] (14) at (-2.25, 3) {};
		\node [style=none] (15) at (-2.25, -0.25) {};
		\node [style=none] (16) at (-2.25, 3) {$A$};
		\node [style=none] (17) at (-2.25, -0.25) {$B$};
		\node [style=none] (18) at (-0.25, 3.5) {};
		\node [style=none] (19) at (-0.25, 2.75) {};
		\node [style=none] (20) at (0.75, 3.5) {};
		\node [style=none] (21) at (0.75, 2.75) {};
		\node [style=none] (23) at (0.25, 2.75) {};
		\node [style=none] (26) at (-0.25, 1.75) {};
		\node [style=none] (27) at (-0.25, 1) {};
		\node [style=none] (28) at (0.75, 1.75) {};
		\node [style=none] (29) at (0.75, 1) {};
		\node [style=none] (31) at (0.25, 1) {};
		\node [style=none] (32) at (0.25, 5) {};
		\node [style=none] (33) at (0.25, 3.5) {};
		\node [style=none] (34) at (0.25, 1.75) {};
		\node [style=none] (35) at (0.25, -0.25) {};
		\node [style=none] (36) at (0.25, 1.5) {};
		\node [style=none] (37) at (0.25, 1.5) {};
		\node [style=none] (38) at (0.25, 1.25) {$g$};
		\node [style=none] (39) at (0.5, 5) {};
		\node [style=none] (40) at (0.5, 2.25) {};
		\node [style=none] (41) at (0.5, -0.25) {};
		\node [style=none] (42) at (0.5, 5) {$A$};
		\node [style=none] (43) at (0.5, 2.25) {$B$};
		\node [style=none] (44) at (0.5, -0.25) {$C$};
		\node [style=none] (45) at (2.75, 1.75) {};
		\node [style=none] (46) at (2.75, 1) {};
		\node [style=none] (47) at (3.75, 1.75) {};
		\node [style=none] (48) at (3.75, 1) {};
		\node [style=none] (49) at (3.25, 1.75) {};
		\node [style=none] (50) at (3.25, 1) {};
		\node [style=none] (51) at (3.25, -0.25) {};
		\node [style=none] (52) at (3.25, 3) {};
		\node [style=none] (53) at (3.25, 1.5) {};
		\node [style=none] (54) at (3.25, 1.25) {$f$};
		\node [style=none] (55) at (3.5, 3) {};
		\node [style=none] (56) at (3.5, -0.25) {};
		\node [style=none] (57) at (3.5, 3) {$A$};
		\node [style=none] (58) at (3.5, -0.25) {$B$};
		\node [style=none] (59) at (4.25, 1.75) {};
		\node [style=none] (60) at (4.25, 1) {};
		\node [style=none] (61) at (5.25, 1.75) {};
		\node [style=none] (62) at (5.25, 1) {};
		\node [style=none] (63) at (4.75, 1.75) {};
		\node [style=none] (64) at (4.75, 1) {};
		\node [style=none] (65) at (4.75, -0.25) {};
		\node [style=none] (66) at (4.75, 3) {};
		\node [style=none] (67) at (4.75, 1.5) {};
		\node [style=none] (68) at (4.75, 1.25) {$g$};
		\node [style=none] (69) at (5, 3) {};
		\node [style=none] (70) at (5, -0.25) {};
		\node [style=none] (71) at (5, 3) {$B$};
		\node [style=none] (72) at (5, -0.25) {$C$};
		\node [style=none] (73) at (0.25, 3) {$f$};
	\end{pgfonlayer}
	\begin{pgfonlayer}{edgelayer}
		\draw (0.center) to (1.center);
		\draw (4.center) to (5.center);
		\draw (4.center) to (6.center);
		\draw (6.center) to (7.center);
		\draw (7.center) to (9.center);
		\draw (9.center) to (10.center);
		\draw (11.center) to (8.center);
		\draw (5.center) to (9.center);
		\draw (18.center) to (19.center);
		\draw (18.center) to (20.center);
		\draw (20.center) to (21.center);
		\draw (21.center) to (23.center);
		\draw (19.center) to (23.center);
		\draw (26.center) to (27.center);
		\draw (28.center) to (29.center);
		\draw (29.center) to (31.center);
		\draw (27.center) to (31.center);
		\draw (26.center) to (28.center);
		\draw (32.center) to (33.center);
		\draw (31.center) to (35.center);
		\draw (45.center) to (46.center);
		\draw (45.center) to (47.center);
		\draw (47.center) to (48.center);
		\draw (48.center) to (50.center);
		\draw (50.center) to (51.center);
		\draw (52.center) to (49.center);
		\draw (46.center) to (50.center);
		\draw (59.center) to (60.center);
		\draw (59.center) to (61.center);
		\draw (61.center) to (62.center);
		\draw (62.center) to (64.center);
		\draw (64.center) to (65.center);
		\draw (66.center) to (63.center);
		\draw (60.center) to (64.center);
		\draw (23.center) to (34.center);
	\end{pgfonlayer}
\end{tikzpicture}
\end{center}

States and effects of an object $A$ are defined as follows, respectively from left to right:
$$ \psi : I \rightarrow A \ \ \ \ \pi : A \rightarrow I $$

\begin{center}
\begin{tikzpicture}
	\begin{pgfonlayer}{nodelayer}
		\node [style=none] (0) at (-5.5, 26.5) {};
		\node [style=none] (1) at (-4, 26.5) {};
		\node [style=none] (2) at (-4.75, 27.5) {};
		\node [style=none] (3) at (-4.75, 26.5) {};
		\node [style=none] (4) at (-4.75, 25.5) {};
		\node [style=none] (5) at (-3, 26.5) {};
		\node [style=none] (6) at (-1.5, 26.5) {};
		\node [style=none] (8) at (-2.25, 26.5) {};
		\node [style=none] (9) at (-2.25, 25.5) {};
		\node [style=none] (10) at (-2.25, 27.5) {};
		\node [style=none] (11) at (-2.25, 27.5) {};
		\node [style=none] (12) at (-4.75, 27) {};
		\node [style=none] (13) at (-4.75, 27) {$\psi$};
		\node [style=none] (14) at (-2.25, 26) {};
		\node [style=none] (15) at (-2.25, 26) {$\pi$};
	\end{pgfonlayer}
	\begin{pgfonlayer}{edgelayer}
		\draw (2.center) to (0.center);
		\draw (2.center) to (1.center);
		\draw (0.center) to (3.center);
		\draw (3.center) to (1.center);
		\draw (3.center) to (4.center);
		\draw (5.center) to (8.center);
		\draw (8.center) to (6.center);
		\draw (5.center) to (9.center);
		\draw (6.center) to (9.center);
		\draw (11.center) to (8.center);
	\end{pgfonlayer}
\end{tikzpicture}
\end{center}

\begin{definition}	
A compact closed category is a monoidal category where for each object $A$ there are objects $A^r$ and $A^l$, and morphisms
$$ \eta ^l : I \rightarrow A\otimes A^l,\ \ \ \ 
\eta ^r : I \rightarrow A^r \otimes A, \ \ \ \ 
 \epsilon^l : A^l \otimes A \rightarrow I, \ \ \ \
\epsilon^r : A \otimes A^r \rightarrow I$$
such that:
\begin{itemize}
\item
 $(1_A \otimes \epsilon^l) \circ (\eta^l \otimes 1_A) = 1_A $
\item
$(\epsilon ^ r \otimes 1_A) \circ (1_A \otimes \eta ^r)=1_A $ 
\item
$(\epsilon^l \otimes 1_{A^l}) \circ (1_{A^l} \otimes \eta^l)=1_{A^l} $ 
\item
$(1_{A^r}\otimes \epsilon^r) \circ (\eta^r \otimes 1_{A^r})=1_{A^r}.$
\end{itemize}
 \end{definition}
 The above equations are called yanking equations. In the graphical language the $\eta$ maps are depicted by caps, and $\epsilon $ maps are depicted by cups \cite{math}. The yanking equation results in a straight wire .
 For example, the diagrams for  $\eta ^l : I \rightarrow A\otimes A^l $, $ \epsilon^l : A^l \otimes A \rightarrow I $ and $(\epsilon^l \otimes 1_{A^l}) \circ (1_{A^l} \otimes \eta^l)=1_{A^l}$ are as follows, respectively from left to right:
\begin{center}
\begin{tikzpicture}
	\begin{pgfonlayer}{nodelayer}
		\node [style=none] (3) at (-1.75, 2) {};
		\node [style=none] (4) at (-3.25, 2) {};
		\node [style=none] (5) at (-6.5, 2) {};
		\node [style=none] (6) at (-5, 2) {};
		\node [style=none] (7) at (1.5, 2) {};
		\node [style=none] (8) at (0, 2) {};
		\node [style=none] (9) at (1.5, 2) {};
		\node [style=none] (10) at (3, 2) {};
		\node [style=none] (11) at (0, 2.75) {};
		\node [style=none] (12) at (3, 1.25) {};
		\node [style=none] (13) at (3.5, 2) {};
		\node [style=none] (14) at (3.5, 2) {$=$};
		\node [style=none] (15) at (4, 3) {};
		\node [style=none] (16) at (4, 1.25) {};
		\node [style=none] (19) at (-1.75, 2.5) {};
		\node [style=none] (20) at (-1.75, 2.5) {$A$};
		\node [style=none] (21) at (-6.5, 1.5) {};
		\node [style=none] (22) at (-6.5, 1.5) {$A$};
		\node [style=none] (23) at (-5, 1.5) {};
		\node [style=none] (24) at (-5, 1.5) {$A^l$};
		\node [style=none] (25) at (-3.25, 2.5) {};
		\node [style=none] (26) at (-3.25, 2.5) {$A^l$};
		\node [style=none] (28) at (4.25, 1.25) {$A^l$};
	\end{pgfonlayer}
	\begin{pgfonlayer}{edgelayer}
		\draw [bend right=90, looseness=1.50] (4.center) to (3.center);
		\draw [bend left=90, looseness=1.25] (5.center) to (6.center);
		\draw [bend right=90, looseness=1.50] (8.center) to (7.center);
		\draw [bend left=90, looseness=1.25] (9.center) to (10.center);
		\draw (8.center) to (11.center);
		\draw (10.center) to (12.center);
		\draw (15.center) to (16.center);
	\end{pgfonlayer}
\end{tikzpicture}

\end{center}

\begin{definition}
As defined in \cite{math}, a partially ordered non-commutative monoid $P$ is called a pregroup, to which we refer as $Preg$. 
Each element $p \in P$ has both a left adjoint $p^l\in P$ and a right adjoint $p^r \in P$. A partially ordered monoid is a set $(P, . ~ , 1,\leq, (-)^l, (-)^r)$ with a partial order relation on $P$ and a binary operation $- \cdot -: P \times P \rightarrow P$ that preserves the partial order relation.
The multiplication has the unit $1$,
that is $ p = 1.p = p.1 $. Explicitly we have the following axioms:
$$ \epsilon_p^l = p^l . p \leq 1, \ \ \ \
 \epsilon_p^r = p . p^r \leq 1, \ \ \ \
 \eta_p^l = 1 \leq p . p^l, \ \ \ \    
 \eta_p^r = 1 \leq p^r . p $$
 We refer the above axioms as reductions.
\end{definition}
\subsection{$\bf{Preg}$ and $\bf{FVect}$ as compact closed categories}

$Preg$ is a compact closed category. Morphisms are  reductions and the operation ‘‘ . ’’ is the monoidal tensor of the monoidal category.
As mentioned in \cite{math},  the category $Preg$ can be used to encoding the grammatical structure of a sentence in a language. 
Objects and morphisms are grammatical types and grammatical reductions, respectively. The operation ‘‘ . ’’ is the juxtaposition of types.
According to \cite{frob}, let $FVect$ be the category of finite dimensional vector spaces over the field of reals $\mathbb{R}$. $FVect$ is a monoidal category, in which vector spaces, linear maps and the tensor product are as objects, morphisms and the monoidal tensor, respectively. In this category the tensor product is commutative, i.e. $ V \otimes W \cong W \otimes V $, and hence $V^l \cong V^r \cong V^*$, where $V^l$, $V^r$ and $V^*$ are left adjoint, right adjoint and a dual space of $V$. 
We consider a fixed base, so we have an inner-product. Consider a vector space $V$ with base $\{\overrightarrow{e_i}\}_i$. Since $V$ is an inner product space with finite dimension, $V^* \cong V$. Therefore $V^r \cong V^l \cong V$,
           \begin{align*}
             \eta^l = \eta^r : 
           &\mathbb{R}\rightarrow V \otimes V \\
           & 1 \mapsto \sum_i \overrightarrow{e_i} \otimes \overrightarrow{e_i}
	\end{align*}
and
          \begin{align*}
          &\epsilon^l=\epsilon^r:
          V \otimes V \rightarrow \mathbb{R} \\
          & \sum_{ij} c_{ij} \overrightarrow{v_i} \otimes \overrightarrow{w_j} 
          \mapsto \sum_{ij} c_{ij} \braket{\overrightarrow{v_i} | \overrightarrow{w_j}}.
\end{align*}
Consider the monoidal functor
$ F : Preg \rightarrow FVect,$
which assigns the basic types to vector spaces as follows:
$$ F(n)=N \ \ \ F(s)=S \ \ \  F(1)=I,$$
and also $F(x \otimes y)=F(x)\otimes F(y)$. 
The compact structure is preserved by Monoidal functors; this means that
$$F(x^r)=F(x^l)=F(x)^*$$ 
for more details see \cite{frob}.

\section{Positive transitive Sentence}	
The simple declarative Persian sentence with a transitive verb has the following structure:

subject + object + objective sign + transitive verb.
For example, the following is the Persian sentence for ‘sara bought the book’.
\begin{center}
Persian: Sara ketab ra kharid.
\\
English: Sara bought the book.
\end{center}

In this sentence, ‘Sara’ is the subject, ‘ketab’ is the direct object, ‘ra’ is the objective sign and ‘kharid’ is the transitive verb in simple past tense, see \cite{m}.
\subsection{Vector Space Interpretation}
Vector spaces and pregroups are used to assign meanings to words and grammatical structure to sentences in a language. The reductions and types are interpreted as linear maps and vector spaces, obtained by a monoidal functor $F$ from $Preg$ to $FVect$.
In this paper we present one example from persian: positive transitive sentence, for which we fix the following basic types, 

n: noun   

s: declarative statement   

o: object

According to \cite{math} if the juxtaposition of the types of the words in a sentence reduces to the basic type s, the sentence is called grammatical. We use an arrow $\rightarrow$ for $~\leq~$ and drop the ‘‘ . ’’ between juxtaposed types. The example sentence ‘sara ketab ra kharid’, has the following type assignment by \cite{m}:

\begin{center}
Sara \ \ \ \ \ ketab \ \ \ \ \ ra \ \ \ \ \ kharid.
\\
$~n\ \ \ \ \ \ \ \ \ \ \  n\ ~~~\ \ \ \ \ \ \ \ (n^r o)\ \ \ \ ( o^r n^r s)$
\end{center}
Which is grammatical because of the following reduction: 
\begin{center}
$n n (n^r o) (o^r n^r s) \rightarrow n11n^r s \rightarrow s$
\end{center}
Reductions are depicted diagrammatically, that of the above is:

\begin{tikzpicture}
	\begin{pgfonlayer}{nodelayer}
		\node [style=none] (0) at (-4.5, 2.75) {};
		\node [style=none] (1) at (-2.25, 2.75) {};
		\node [style=none] (2) at (-0.75, 2.75) {};
		\node [style=none] (3) at (0.25, 2.75) {};
		\node [style=none] (4) at (2, 2.75) {};
		\node [style=none] (5) at (2.75, 2.75) {};
		\node [style=none] (6) at (2.75, 2.75) {};
		\node [style=none] (7) at (3.25, 0.75) {};
		\node [style=none] (8) at (3.25, 2.75) {};
		\node [style=none] (9) at (-4.5, 3.25) {};
		\node [style=none] (11) at (-0.75, 3.25) {};
		\node [style=none] (12) at (0.25, 3.25) {};
		\node [style=none] (13) at (2, 3.25) {};
		\node [style=none] (14) at (2.75, 3.25) {};
		\node [style=none] (15) at (3.25, 3.25) {};
		\node [style=none] (16) at (-4.5, 3.25) {$n$};
		\node [style=none] (17) at (-2.25, 3.25) {};
		\node [style=none] (18) at (-2.25, 3.25) {$n$};
		\node [style=none] (19) at (-0.75, 3.25) {$n^r$};
		\node [style=none] (20) at (0.25, 3.25) {$o$};
		\node [style=none] (21) at (2, 3.25) {$o^r$};
		\node [style=none] (22) at (2.75, 3.25) {$n^r$};
		\node [style=none] (23) at (3.25, 3.25) {$s$};
		\node [style=none] (24) at (5.75, 2.5) {$(1)$};
	\end{pgfonlayer}
	\begin{pgfonlayer}{edgelayer}
		\draw [bend right=90, looseness=2.00] (1.center) to (2.center);
		\draw [bend right=90, looseness=1.75] (3.center) to (4.center);
		\draw [bend right=90] (0.center) to (6.center);
		\draw (8.center) to (7.center);
	\end{pgfonlayer}
\end{tikzpicture}
 
A positive sentence with a transitive verb in Persian has the pregroup type $ n n (n^r o) (o^r n^r s)$.
The interpretation of a transitive verb is computed as follows:
$$F(o^r\otimes n^r \otimes s)=F(o^r)\otimes F(n^r)\otimes F(s)=F(o)^r\otimes F(n)^r\otimes F(s)=
$$
$$
F(o)^*\otimes F(n)^*\otimes F(s)=N\otimes N\otimes S $$
So the meaning vector of a Persian transitive verb is a vector in $N\otimes N\otimes S.$
The pregroup reduction of a  transitive sentence is computed as follows:
$$ F ((\epsilon_n^r \otimes 1_s) \circ (1_n \otimes \epsilon_n^r \otimes \epsilon_o^r \otimes 1_n \otimes 1_s)) =$$
$$ (F(\epsilon_n^r)\otimes F(1_s)) \circ 
(F(1_n)\otimes F(\epsilon_n^r) \otimes F(\epsilon_o^r)\otimes F(1_s)=$$ 
$$(F(\epsilon_n)^* \otimes F(1_s)) \circ 
(F(1_n)\otimes F(\epsilon_n)^* \otimes F(\epsilon_o)^* \otimes F(1_s)=$$
$$(\epsilon_N \otimes 1_S) \circ (1_N \otimes \epsilon_N \otimes \epsilon_N \otimes 1_N \otimes 1_S)$$
 and depicted as:
 \vspace{1cm}
  \begin{center}
\begin{tikzpicture}
	\begin{pgfonlayer}{nodelayer}
		\node [style=none] (1) at (-3.25, 3.5) {};
		\node [style=none] (2) at (-1.75, 3.5) {};
		\node [style=none] (3) at (-1.25, 3.5) {};
		\node [style=none] (4) at (0.5, 3.5) {};
		\node [style=none] (5) at (1.25, 3.5) {};
		\node [style=none] (6) at (1.75, 3.5) {};
		\node [style=none] (7) at (1.75, 1.75) {};
		\node [style=none] (8) at (-4.75, 3.5) {};
		\node [style=none] (9) at (-4.75, 2.75) {};
		\node [style=none] (10) at (1.25, 2.75) {};
	\end{pgfonlayer}
	\begin{pgfonlayer}{edgelayer}
		\draw [bend right=90, looseness=1.50] (1.center) to (2.center);
		\draw [bend right=90, looseness=1.25] (3.center) to (4.center);
		\draw (6.center) to (7.center);
		\draw (8.center) to (9.center);
		\draw (5.center) to (10.center);
		\draw [bend right=90, looseness=0.50] (9.center) to (10.center);
	\end{pgfonlayer}
\end{tikzpicture}
\end{center}
The distributional meaning of ‘Sara ketab ra kharid’
is as follows:
$$F ((\epsilon_n^r \otimes 1_s) \circ (1_n \otimes \epsilon_n^r \otimes \epsilon_o^r \otimes 1_n \otimes 1_s)) (\overrightarrow{sara} \otimes \overrightarrow{ketab} \otimes \overrightarrow{ra} \otimes \overrightarrow{kharid})$$

where $\overrightarrow{ra}$ is the vector corresponding to the meaning of ‘ra’. We set 
$$\overrightarrow{ra} = \sum_i \overrightarrow{e_i} \otimes \overrightarrow{e_i} \in N \otimes N $$
and in this case we have
$$\overrightarrow{ra} \simeq \eta_N : \mathbf{R}\rightarrow N \otimes N. $$

We obtain diagrammatically:
\vspace{1cm}
\begin{center}
\begin{tikzpicture}[thick]
	\begin{pgfonlayer}{nodelayer}
		\node [style=none] (3) at (-4, 2) {};
		\node [style=none] (4) at (-5.5, 2) {};
		\node [style=none] (5) at (-4, 2) {};
		\node [style=none] (6) at (-2.5, 2) {};
		\node [style=none] (7) at (-1, 2) {};
		\node [style=none] (8) at (-2.5, 2) {};
		\node [style=none] (9) at (-1, 2) {};
		\node [style=none] (10) at (-6, 2) {};
		\node [style=none] (11) at (-5, 2) {};
		\node [style=none] (12) at (-5.5, 2.75) {};
		\node [style=none] (13) at (-1.5, 2) {};
		\node [style=none] (14) at (0, 2) {};
		\node [style=none] (15) at (-0.75, 2.75) {};
		\node [style=none] (16) at (-0.75, 1) {};
		\node [style=none] (17) at (-0.25, 0.25) {};
		\node [style=none] (18) at (-0.75, 2) {};
		\node [style=none] (19) at (-0.25, 2) {};
		\node [style=none] (20) at (-6.5, 2) {};
		\node [style=none] (21) at (-7, 2) {};
		\node [style=none] (22) at (-7.5, 2) {};
		\node [style=none] (23) at (-7, 2.75) {};
		\node [style=none] (24) at (-7, 1) {};
		\node [style=none] (25) at (-7, 3.25) {$subject$};
		\node [style=none] (26) at (-5.5, 3.25) {$object$};
		\node [style=none] (27) at (-0.75, 3.25) {$verb$};
	\end{pgfonlayer}
	\begin{pgfonlayer}{edgelayer}[thick]
		\draw [bend right=90, looseness=2.00] (4.center) to (3.center);
		\draw [bend left=90, looseness=1.50] (5.center) to (6.center);
		\draw [bend right=90, looseness=1.75] (8.center) to (7.center);
		\draw (10.center) to (11.center);
		\draw (12.center) to (10.center);
		\draw (12.center) to (11.center);
		\draw (13.center) to (9.center);
		\draw (9.center) to (14.center);
		\draw (15.center) to (13.center);
		\draw (15.center) to (14.center);
		\draw (19.center) to (17.center);
		\draw (18.center) to (16.center);
		\draw (23.center) to (22.center);
		\draw (22.center) to (21.center);
		\draw (21.center) to (20.center);
		\draw (23.center) to (20.center);
		\draw (21.center) to (24.center);
		\draw [bend right=90, looseness=0.50] (24.center) to (16.center);
	\end{pgfonlayer}
\end{tikzpicture}
\end{center}

Which by the diagrammatic calculus of compact closed categories \cite{picturing}, is equal to:

 \begin{center}
  \begin{tikzpicture}
  	\begin{pgfonlayer}{nodelayer}
  		\node [style=none] (4) at (-3, 2) {};
  		\node [style=none] (7) at (-1, 2) {};
  		\node [style=none] (9) at (-1, 2) {};
  		\node [style=none] (10) at (-3.5, 2) {};
  		\node [style=none] (11) at (-2.5, 2) {};
  		\node [style=none] (12) at (-3, 2.75) {};
  		\node [style=none] (13) at (-1.5, 2) {};
  		\node [style=none] (14) at (0, 2) {};
  		\node [style=none] (15) at (-0.75, 2.75) {};
  		\node [style=none] (17) at (-0.25, 0.25) {};
  		\node [style=none] (18) at (-1, 2) {};
  		\node [style=none] (19) at (-0.25, 2) {};
  		\node [style=none] (20) at (-6.5, 2) {};
  		\node [style=none] (21) at (-7, 2) {};
  		\node [style=none] (22) at (-7.5, 2) {};
  		\node [style=none] (23) at (-7, 2.75) {};
  		\node [style=none] (24) at (-7, 1) {};
  		\node [style=none] (25) at (-0.75, 2) {};
  		\node [style=none] (26) at (-0.75, 2) {};
  		\node [style=none] (27) at (-0.75, 1) {};
  		\node [style=none] (28) at (-7, 3.25) {$subject$};
  		\node [style=none] (29) at (-3, 3.25) {$object$};
  		\node [style=none] (30) at (-0.75, 3.25) {$verb$};
  	\end{pgfonlayer}
  	\begin{pgfonlayer}{edgelayer}[thick]
  		\draw (10.center) to (11.center);
  		\draw (12.center) to (10.center);
  		\draw (12.center) to (11.center);
  		\draw (13.center) to (9.center);
  		\draw (9.center) to (14.center);
  		\draw (15.center) to (13.center);
  		\draw (15.center) to (14.center);
  		\draw (19.center) to (17.center);
  		\draw (23.center) to (22.center);
  		\draw (22.center) to (21.center);
  		\draw (21.center) to (20.center);
  		\draw (23.center) to (20.center);
  		\draw (21.center) to (24.center);
  		\draw [bend right=90, looseness=1.50] (4.center) to (18.center);
  		\draw (26.center) to (27.center);
  		\draw [bend right=90, looseness=0.50] (24.center) to (27.center);
  	\end{pgfonlayer}
  \end{tikzpicture}
  \end{center}
Consider the vector $\Psi$ in the tensor space which represents the type of verb:
$$\Psi=\sum_{ijk} c_{ijk} ~~\overrightarrow{w_i} \otimes \overrightarrow{v_j} \otimes \overrightarrow{s_k}\in N \otimes N \otimes S $$
where for each $i$, $\overrightarrow{w_i}$ is the meaning vector of object and $\overrightarrow{v_j}$ is the meaning vector of subject. Then
$$(\epsilon_N \otimes 1_S) \circ (1_N \otimes \epsilon_N \otimes 1_N \otimes 1_S)(\overrightarrow{v} \otimes \overrightarrow{w} \otimes \overrightarrow{\Psi})= $$
$$(\epsilon_N \otimes 1_S) \circ (1_N \otimes \epsilon_N \otimes 1_N \otimes 1_S)(\overrightarrow{v} \otimes \overrightarrow{w} \otimes (\sum_{ijk} c_{ijk} ~~ \overrightarrow{w_i} \otimes \overrightarrow{v_j} \otimes \overrightarrow{s_k}))=$$
$$(\epsilon_N \otimes 1_S)(\overrightarrow{v} \otimes \braket{w,w_i} \otimes \overrightarrow{v_j}\otimes \overrightarrow{s_k})=
\sum_{ijk} c_{ijk} ~~ \braket{w,w_i} \braket{v,v_j}\overrightarrow{s_k}.$$
\subsection{Truth theoretic meaning and concrete instantiation}

According to \cite{frob} we let $N$ to be the vector space spanned by a set of individuals $\{\overrightarrow{n_i} \}$ and $S$ to be the one dimensional space spanned by the unit vector $\overrightarrow{1}$. The unit vector and the zero vector represent truth value $1$ and truth value $0$ respectively. A transitive verb $\varPsi \in N\otimes N \otimes S$ is represented as follows:

$$\varPsi:= \sum_{ji} \overrightarrow{n_j}\otimes \overrightarrow{n_i}\otimes (\alpha_{ji}\overrightarrow{1}) $$
where $\overrightarrow{sub}=\sum_i \overrightarrow{n_i}$, $~~ \overrightarrow{obj}=\sum_j \overrightarrow{n_j}$ and $\alpha_{ji}$'s are degrees of truth, i.e. $\overrightarrow{n_i}$  $\varPsi$'s $\overrightarrow{n_j}$ with degree $\alpha_{ji}$, for all $i,j$. 
For $\overrightarrow{sub}=\sum_k \overrightarrow{n_k}$ and $\overrightarrow{obj}=\sum_l \overrightarrow{n_l}$, where $k$ and $l$ range  over the  sets of basis  vectors  representing  the respective  common nouns, the truth-theoretic meaning of a transitive sentence is computed as follows:
$$\overrightarrow{sub \ \ obj \ \ ra \ \ verb}= (\epsilon_N \otimes 1_S) \circ (1_N \otimes \epsilon_N \otimes 1_N \otimes 1_S)(\overrightarrow{sub} \otimes \overrightarrow{obj} \otimes \overrightarrow{verb})$$
$$=(\epsilon_N \otimes 1_S) \circ (1_N \otimes \epsilon_N \otimes 1_N \otimes 1_S)(\sum_k \overrightarrow{n_k} \otimes \sum_l \overrightarrow{n_l} \otimes (\sum_{ji}\overrightarrow{n_j} \otimes \overrightarrow{n_i} \otimes \alpha_{ji}\overrightarrow{1}))$$
$$=\sum_{kl}\alpha_{kl}\overrightarrow{1}.$$

For concrete instantiation in the model of Grefenstette and Sadrzadeh \cite{e} 
the vectors are obtained from corpora and the scalar weights for noun vectors are not necessarily $1$ or $0$.
For any word vector $\overrightarrow{word}=\sum c_i ^{word} \overrightarrow{n_i}$, the scalar weight $c_i ^{word}$
is the number  of  times that the word has appeared in that context. Where $\overrightarrow{n_i}$'s are context basis vectors.
The meaning of the transitive sentence is:
$$\overrightarrow{sub \ \ obj \ \ ra \ \ verb}= \sum_{jit}\braket{\overrightarrow{obj} | \overrightarrow{n_j}} \braket{\overrightarrow{sub} | \overrightarrow{n_i}} c_{jit} \overrightarrow{s_t} $$
A transitive verb is represented as a two dimensional matrix. The corresponding vector of this matrix is 
$\overrightarrow{verb}=\sum_{ji}c_{ji}(\overrightarrow{n_j}\otimes \overrightarrow{n_i}).$
Note that the sum of the tensor product of the objects and subjects of the verb throughout a corpus represents the meaning vector of the verb.
So the meaning of the transitive sentence is:
        
         \begin{align*}
          \overrightarrow{sub \ \ obj \ \ ra \ \ verb} &=\sum_{ji}\braket{\overrightarrow{obj} | \overrightarrow{n_j}} \braket{\overrightarrow{sub} | \overrightarrow{n_i}} c_{ji}(\overrightarrow{n_j}\otimes \overrightarrow{n_i})\\
         & =\sum_{ji} c_j ^{obj} c_i ^{sub} c_{ji}(\overrightarrow{n_j}\otimes \overrightarrow{n_i}).
         \end{align*}
     
  The meaning vector is decomposed to point-wise multiplication of two vectors as follows:
 $$(\sum_{ji} c_j ^{obj} c_i ^{sub} (\overrightarrow{n_j}\otimes \overrightarrow{n_i}))\odot (\sum_{ji} c_{ji} (\overrightarrow{n_j}\otimes \overrightarrow{n_i}))$$
 $$=(\overrightarrow{obj} \otimes \overrightarrow{sub}) \odot \overrightarrow{verb}$$
where $\odot$ is the point-wise multiplication.

\section{Diagrams rewriting and quantum circuits}
As mentioned in the previous sections a sentence in a corpus is parsed according to its grammatical structure. According to \cite{circuit} we simplify the DisCoCat diagram to some other diagram and turn into a quantum circuit, which can be compiled via NISQ devices. Two methods are presented for this purpose. The bigraph method and snake removal method. Both methods are done in the symmetric version of the pregroup grammar. 
We consider the grammatical sentence from (1), 
\vspace{1cm}
\begin{center}
\begin{tikzpicture}
	\begin{pgfonlayer}{nodelayer}
		\node [style=none] (0) at (-5.75, 2.5) {};
		\node [style=none] (1) at (-3, 2.5) {};
		\node [style=none] (2) at (-0.75, 2.5) {};
		\node [style=none] (3) at (0.25, 2.5) {};
		\node [style=none] (4) at (2.25, 2.5) {};
		\node [style=none] (5) at (3.75, 2.5) {};
		\node [style=none] (6) at (3.25, 2.5) {};
		\node [style=none] (7) at (3.75, 0.25) {};
		\node [style=none] (8) at (3.75, 2.5) {};
		\node [style=none] (9) at (-6.75, 2.5) {};
		\node [style=none] (10) at (-4.75, 2.5) {};
		\node [style=none] (14) at (-4, 2.5) {};
		\node [style=none] (15) at (-2, 2.5) {};
		\node [style=none] (18) at (-3, 3.25) {};
		\node [style=none] (19) at (-1.25, 2.5) {};
		\node [style=none] (20) at (0.75, 2.5) {};
		\node [style=none] (24) at (-0.25, 3) {};
		\node [style=none] (26) at (-0.25, 3.25) {};
		\node [style=none] (28) at (1.75, 2.5) {};
		\node [style=none] (29) at (4.5, 2.5) {};
		\node [style=none] (32) at (3.25, 3.25) {};
		\node [style=none] (33) at (-5.75, 3.25) {};
		\node [style=none] (34) at (-5.75, 2.75) {$Sara$};
		\node [style=none] (35) at (-3, 2.75) {};
		\node [style=none] (36) at (-3, 2.75) {$ketab$};
		\node [style=none] (37) at (-0.25, 2.75) {};
		\node [style=none] (38) at (-0.25, 2.75) {$ra$};
		\node [style=none] (39) at (3.25, 2.75) {};
		\node [style=none] (40) at (3.25, 2.75) {$kharid$};
		\node [style=none] (41) at (-6.75, 3) {};
		\node [style=none] (42) at (-4.75, 3) {};
		\node [style=none] (43) at (-4, 3) {};
		\node [style=none] (44) at (-2, 3) {};
		\node [style=none] (45) at (-1.25, 3) {};
		\node [style=none] (46) at (0.75, 3) {};
		\node [style=none] (47) at (1.75, 3) {};
		\node [style=none] (48) at (4.5, 3) {};
		\node [style=none] (49) at (5.75, 2.5) {$(2)$};
	\end{pgfonlayer}
	\begin{pgfonlayer}{edgelayer}
		\draw [bend right=90, looseness=1.50] (1.center) to (2.center);
		\draw [bend right=90, looseness=1.25] (3.center) to (4.center);
		\draw [bend right=90, looseness=0.75] (0.center) to (6.center);
		\draw (8.center) to (7.center);
		\draw (9.center) to (0.center);
		\draw (0.center) to (10.center);
		\draw (14.center) to (1.center);
		\draw (15.center) to (1.center);
		\draw (19.center) to (2.center);
		\draw (20.center) to (3.center);
		\draw (3.center) to (2.center);
		\draw (28.center) to (4.center);
		\draw (4.center) to (6.center);
		\draw (6.center) to (8.center);
		\draw (8.center) to (29.center);
		\draw (41.center) to (9.center);
		\draw (42.center) to (10.center);
		\draw (41.center) to (33.center);
		\draw (33.center) to (42.center);
		\draw (43.center) to (14.center);
		\draw (44.center) to (15.center);
		\draw (18.center) to (43.center);
		\draw (18.center) to (44.center);
		\draw (45.center) to (19.center);
		\draw (46.center) to (20.center);
		\draw (45.center) to (26.center);
		\draw (26.center) to (46.center);
		\draw (47.center) to (28.center);
		\draw (32.center) to (47.center);
		\draw (48.center) to (29.center);
		\draw (48.center) to (32.center);
	\end{pgfonlayer}
\end{tikzpicture}
\end{center}
and use a bigraph method to turn the diagram
(2) into a bipartite graph. Words at odd distance from the root word are transposed into effects:
\begin{center}
\begin{tikzpicture}
	\begin{pgfonlayer}{nodelayer}
		\node [style=none] (11) at (2.75, 2.5) {};
		\node [style=none] (12) at (1.75, 2.5) {};
		\node [style=none] (13) at (2.75, 2.5) {};
		\node [style=none] (14) at (0.75, 2.5) {};
		\node [style=none] (15) at (3.5, 2.5) {};
		\node [style=none] (17) at (2.25, 3.25) {};
		\node [style=none] (18) at (2.25, 2.75) {};
		\node [style=none] (19) at (2.25, 2.75) {$kharid$};
		\node [style=none] (20) at (0.75, 3) {};
		\node [style=none] (21) at (3.5, 3) {};
		\node [style=none] (22) at (-3.75, 2.5) {};
		\node [style=none] (23) at (-4.75, 2.5) {};
		\node [style=none] (24) at (-2.75, 2.5) {};
		\node [style=none] (25) at (-3.75, 3.25) {};
		\node [style=none] (26) at (-3.75, 2.75) {};
		\node [style=none] (27) at (-3.75, 2.75) {$ketab$};
		\node [style=none] (28) at (-4.75, 3) {};
		\node [style=none] (29) at (-2.75, 3) {};
		\node [style=none] (31) at (-0.75, 0.75) {};
		\node [style=none] (33) at (-0.75, 0.25) {};
		\node [style=none] (34) at (-1.5, 0.25) {};
		\node [style=none] (35) at (-1.5, 0) {};
		\node [style=none] (36) at (-1.5, 0.5) {$ra$};
		\node [style=none] (37) at (1, 0.75) {};
		\node [style=none] (38) at (2.5, 0.75) {};
		\node [style=none] (39) at (1, 0.25) {};
		\node [style=none] (40) at (2.5, 0.25) {};
		\node [style=none] (41) at (1.75, 0) {};
		\node [style=none] (42) at (1.75, 0.5) {$Sara$};
		\node [style=none] (45) at (2.75, 0.5) {};
		\node [style=none] (46) at (1.75, 0.75) {};
		\node [style=none] (47) at (2, 1.25) {};
		\node [style=none] (49) at (-1.25, 0.75) {};
		\node [style=none] (50) at (1.25, 2.5) {};
		\node [style=none] (55) at (-2.25, 0.75) {};
		\node [style=none] (56) at (-2.25, 0.25) {};
		\node [style=none] (57) at (-1.75, 0.75) {};
		\node [style=none] (58) at (5.75, 2.25) {$(3)$};
	\end{pgfonlayer}
	\begin{pgfonlayer}{edgelayer}
		\draw (12.center) to (13.center);
		\draw (13.center) to (15.center);
		\draw (20.center) to (14.center);
		\draw (17.center) to (20.center);
		\draw (23.center) to (22.center);
		\draw (24.center) to (22.center);
		\draw (28.center) to (23.center);
		\draw (29.center) to (24.center);
		\draw (25.center) to (28.center);
		\draw (25.center) to (29.center);
		\draw (31.center) to (33.center);
		\draw (35.center) to (33.center);
		\draw (37.center) to (39.center);
		\draw (38.center) to (37.center);
		\draw (38.center) to (40.center);
		\draw (39.center) to (41.center);
		\draw (40.center) to (41.center);
		\draw (13.center) to (45.center);
		\draw (12.center) to (46.center);
		\draw [in=90, out=-75] (22.center) to (49.center);
		\draw (14.center) to (12.center);
		\draw (55.center) to (49.center);
		\draw (49.center) to (31.center);
		\draw (55.center) to (56.center);
		\draw (56.center) to (35.center);
		\draw [in=90, out=-90] (50.center) to (57.center);
		\draw (21.center) to (15.center);
		\draw (21.center) to (17.center);
	\end{pgfonlayer}
\end{tikzpicture}
\end{center}
Transposition turns states into effects, see \cite{picturing}. According to \cite{circuit}, we consider CNOT+U(3) of unitary qubit ansatze. Layers of CNOT gates between adjacent qubits with layers of single-qubits rotations in $Z$ and $X$ form unitary quantum circuits. Let $\hat{P}$ be the symmetric version of the pregroup grammar $P$. 
Consider the monoidal functor from $\hat{P}$ to fHilb, in which word states are mapped to the state ansatze. State ansatze are obtained by applying the unitary ansatze to the pauli $Z\ket{0}$ state. Word effects are mapped to the effect ansatze, in which effect ansatze are obtained by transposing the state ansatze in the computational basis, and wire crossing are mapped to swaps.
Now consider the diagram (3). If each wire is mapped to a qubit, the circuit has about four CNOTs.
In ZX-calculus \cite{inter}, suppose single-qubits white and black dots are rotations in pauli Z and pauli X. A CNOT gate is Black and white dots connected by a horizontal line: 

\begin{tikzpicture}
	\begin{pgfonlayer}{nodelayer}
		\node [style=none] (0) at (-8, 0.75) {};
		\node [style=none] (1) at (-8, 0.25) {};
		\node [style=none] (2) at (-6.5, 0.25) {};
		\node [style=none] (3) at (-6.5, 0.75) {};
		\node [style=none] (4) at (-7.25, 1) {};
		\node [style=none] (5) at (-7.25, -1) {};
		\node [style=none] (6) at (-7.25, 0.25) {};
		\node [style=none] (7) at (-7.25, 0.5) {};
		\node [style=none] (8) at (-7.25, 0.5) {$ketab$};
		\node [style=none] (9) at (-5.75, 0.5) {$\mapsto$};
		\node [style=new style 1] (12) at (-4.25, -0.25) {};
		\node [style=none] (13) at (-4.25, -1) {};
		\node [style=none] (14) at (-3.75, 0.5) {};
		\node [style=none] (15) at (-3.75, -0.25) {};
		\node [style=none] (16) at (-3.75, 0.5) {$\theta_0$};
		\node [style=none] (17) at (-3.75, -0.25) {$\theta_1$};
		\node [style=none] (18) at (-8, -5.25) {};
		\node [style=none] (19) at (-8, -4.75) {};
		\node [style=none] (20) at (-6.5, -4.75) {};
		\node [style=none] (21) at (-6.5, -5.25) {};
		\node [style=none] (22) at (-7.25, -5.5) {};
		\node [style=none] (23) at (-7.25, -4.75) {};
		\node [style=none] (24) at (-7.25, -3.25) {};
		\node [style=none] (25) at (-7.25, -5) {};
		\node [style=none] (26) at (-7.25, -5) {$sara$};
		\node [style=none] (27) at (-6, -4) {};
		\node [style=none] (28) at (-6, -4) {$:=$};
		\node [style=none] (29) at (-5.25, -3.75) {};
		\node [style=none] (30) at (-5.25, -4.25) {};
		\node [style=none] (31) at (-3.75, -3.75) {};
		\node [style=none] (32) at (-3.75, -4.25) {};
		\node [style=none] (33) at (-4.5, -3.5) {};
		\node [style=none] (34) at (-4.5, -4.25) {};
		\node [style=none] (35) at (-2.5, -4.25) {};
		\node [style=none] (36) at (-4.5, -4) {$sara$};
		\node [style=none] (37) at (-1.75, -4) {};
		\node [style=none] (38) at (-1.75, -4) {$\mapsto$};
		\node [style=new style 1] (41) at (-0.75, -3.75) {};
		\node [style=none] (42) at (-0.75, -3) {};
		\node [style=none] (44) at (-0.25, -3.75) {};
		\node [style=none] (45) at (-0.25, -4.5) {};
		\node [style=none] (46) at (-0.25, -3.75) {};
		\node [style=none] (47) at (-0.25, -3.75) {};
		\node [style=none] (48) at (-0.25, -3.75) {$\theta_1$};
		\node [style=none] (49) at (-0.25, -4.5) {$\theta_0$};
		\node [style=none] (50) at (-7.5, -12) {};
		\node [style=none] (51) at (-7.5, -12.5) {};
		\node [style=none] (52) at (-6, -12.5) {};
		\node [style=none] (53) at (-6, -12) {};
		\node [style=none] (55) at (-6.75, -12.75) {};
		\node [style=none] (56) at (-7, -12) {};
		\node [style=none] (57) at (-6.5, -12) {};
		\node [style=none] (58) at (-5.25, -10.25) {};
		\node [style=none] (59) at (-8, -10.25) {};
		\node [style=none] (60) at (-6.75, -12.25) {};
		\node [style=none] (61) at (-6.75, -12.25) {};
		\node [style=none] (62) at (-6.75, -12.25) {};
		\node [style=none] (63) at (-6.75, -12.25) {};
		\node [style=none] (64) at (-6.75, -12.25) {$ra$};
		\node [style=none] (65) at (-4.25, -11) {};
		\node [style=none] (66) at (-4.25, -11) {$\mapsto$};
		\node [style=new style 0] (70) at (-3, -13.75) {};
		\node [style=new style 1] (71) at (-3, -12.75) {};
		\node [style=new style 0] (72) at (-3, -11.75) {};
		\node [style=new style 1] (73) at (-3, -10.75) {};
		\node [style=new style 0] (74) at (-3, -9.75) {};
		\node [style=new style 0] (76) at (-1.5, -13.75) {};
		\node [style=new style 1] (77) at (-1.5, -12.75) {};
		\node [style=new style 0] (78) at (-1.5, -11.75) {};
		\node [style=new style 1] (79) at (-1.5, -10.75) {};
		\node [style=new style 0] (80) at (-1.5, -9.75) {};
		\node [style=new style 1] (81) at (-1.5, -13.75) {};
		\node [style=new style 1] (82) at (-1.5, -12.75) {};
		\node [style=new style 0] (83) at (-1.5, -12.75) {};
		\node [style=new style 0] (84) at (-1.5, -11.75) {};
		\node [style=new style 1] (85) at (-1.5, -11.75) {};
		\node [style=new style 0] (86) at (-1.5, -10.75) {};
		\node [style=new style 1] (88) at (-1.5, -9.75) {};
		\node [style=none] (89) at (-1, -13.75) {};
		\node [style=none] (90) at (-1, -12.75) {};
		\node [style=none] (91) at (-1, -10.75) {};
		\node [style=none] (92) at (-1, -9.75) {};
		\node [style=none] (94) at (-2.5, -12.75) {};
		\node [style=none] (95) at (-2.5, -10.75) {};
		\node [style=none] (96) at (-2.5, -9.75) {};
		\node [style=none] (97) at (-2.5, -13.75) {};
		\node [style=none] (98) at (-1, -13.75) {$\theta_0$};
		\node [style=none] (99) at (-1, -12.75) {$\theta_2$};
		\node [style=none] (100) at (-1, -10.75) {$\theta_4$};
		\node [style=none] (101) at (-1, -9.75) {$\theta_6$};
		\node [style=none] (102) at (-2.5, -13.75) {$\theta_1$};
		\node [style=none] (103) at (-2.5, -12.75) {$\theta_3$};
		\node [style=none] (104) at (-2.5, -10.75) {$\theta_5$};
		\node [style=none] (105) at (-2.5, -9.75) {$\theta_7$};
		\node [style=none] (106) at (-1.5, -8.5) {};
		\node [style=none] (107) at (-3, -8.5) {};
		\node [style=none] (108) at (-0.25, -11) {$=$};
		\node [style=new style 0] (109) at (2.25, -14.75) {};
		\node [style=new style 0] (110) at (2.25, -13.75) {};
		\node [style=new style 1] (111) at (2.25, -12.75) {};
		\node [style=new style 0] (112) at (2.25, -11.75) {};
		\node [style=new style 1] (113) at (2.25, -10.75) {};
		\node [style=new style 0] (114) at (2.25, -9.75) {};
		\node [style=none] (115) at (2.75, -12.75) {};
		\node [style=none] (116) at (2.75, -10.75) {};
		\node [style=none] (117) at (2.75, -9.75) {};
		\node [style=none] (118) at (2.75, -13.75) {};
		\node [style=none] (119) at (2.75, -13.75) {$\theta_1$};
		\node [style=none] (120) at (2.75, -12.75) {$\theta_3$};
		\node [style=none] (121) at (2.75, -10.75) {$\theta_5$};
		\node [style=none] (122) at (2.75, -9.75) {$\theta_7$};
		\node [style=new style 0] (125) at (0.75, -14.75) {};
		\node [style=new style 0] (126) at (0.75, -13.75) {};
		\node [style=new style 1] (127) at (0.75, -12.75) {};
		\node [style=new style 0] (128) at (0.75, -11.75) {};
		\node [style=new style 1] (129) at (0.75, -10.75) {};
		\node [style=new style 0] (130) at (0.75, -9.75) {};
		\node [style=new style 1] (131) at (0.75, -13.75) {};
		\node [style=new style 1] (132) at (0.75, -12.75) {};
		\node [style=new style 0] (133) at (0.75, -12.75) {};
		\node [style=new style 0] (134) at (0.75, -11.75) {};
		\node [style=new style 1] (135) at (0.75, -11.75) {};
		\node [style=new style 0] (136) at (0.75, -10.75) {};
		\node [style=new style 1] (137) at (0.75, -9.75) {};
		\node [style=none] (138) at (1.25, -13.75) {};
		\node [style=none] (139) at (1.25, -12.75) {};
		\node [style=none] (140) at (1.25, -10.75) {};
		\node [style=none] (141) at (1.25, -9.75) {};
		\node [style=none] (142) at (1.25, -13.75) {$\theta_0$};
		\node [style=none] (143) at (1.25, -12.75) {$\theta_2$};
		\node [style=none] (144) at (1.25, -10.75) {$\theta_4$};
		\node [style=none] (145) at (1.25, -9.75) {$\theta_6$};
		\node [style=none] (146) at (0.75, -9) {};
		\node [style=none] (147) at (2.25, -9) {};
		\node [style=black dot] (148) at (-4.25, 1.25) {};
		\node [style=black dot] (149) at (-4.25, 0.5) {};
		\node [style=black dot] (150) at (-0.75, -4.5) {};
		\node [style=black dot] (151) at (-0.75, -5.25) {};
		\node [style=black dot] (152) at (-3, -9.75) {};
		\node [style=black dot] (153) at (-1.5, -10.75) {};
		\node [style=black dot] (154) at (-3, -11.75) {};
		\node [style=black dot] (155) at (-1.5, -12.75) {};
		\node [style=black dot] (156) at (-3, -13.75) {};
		\node [style=black dot] (157) at (-3, -14.75) {};
		\node [style=black dot] (158) at (-3, -14.75) {};
		\node [style=black dot] (159) at (-1.5, -14.75) {};
		\node [style=black dot] (160) at (2.25, -9.75) {};
		\node [style=black dot] (161) at (0.75, -10.75) {};
		\node [style=black dot] (162) at (2.25, -11.75) {};
		\node [style=black dot] (163) at (0.75, -12.75) {};
		\node [style=black dot] (164) at (0.75, -14.75) {};
		\node [style=black dot] (165) at (2.25, -13.75) {};
		\node [style=black dot] (166) at (2.25, -14.75) {};
	\end{pgfonlayer}
	\begin{pgfonlayer}{edgelayer}
		\draw (0.center) to (1.center);
		\draw (0.center) to (4.center);
		\draw (4.center) to (3.center);
		\draw (3.center) to (2.center);
		\draw (1.center) to (2.center);
		\draw (6.center) to (5.center);
		\draw (12) to (13.center);
		\draw (19.center) to (18.center);
		\draw (19.center) to (20.center);
		\draw (20.center) to (21.center);
		\draw (18.center) to (22.center);
		\draw (21.center) to (22.center);
		\draw (23.center) to (24.center);
		\draw (29.center) to (30.center);
		\draw (29.center) to (33.center);
		\draw (33.center) to (31.center);
		\draw (31.center) to (32.center);
		\draw (32.center) to (30.center);
		\draw [bend right=90, looseness=1.50] (34.center) to (35.center);
		\draw (42.center) to (41);
		\draw (50.center) to (51.center);
		\draw (50.center) to (53.center);
		\draw (53.center) to (52.center);
		\draw (52.center) to (55.center);
		\draw (51.center) to (55.center);
		\draw [in=-105, out=90] (56.center) to (58.center);
		\draw [in=90, out=-90] (59.center) to (57.center);
		\draw (70) to (71);
		\draw (72) to (71);
		\draw (73) to (72);
		\draw (74) to (73);
		\draw (80) to (79);
		\draw (79) to (78);
		\draw (78) to (77);
		\draw (77) to (76);
		\draw (72) to (85);
		\draw [in=90, out=-90] (106.center) to (74);
		\draw [in=90, out=-90] (107.center) to (88);
		\draw (109) to (110);
		\draw (110) to (111);
		\draw (112) to (111);
		\draw (113) to (112);
		\draw (114) to (113);
		\draw (130) to (129);
		\draw (129) to (128);
		\draw (128) to (127);
		\draw (127) to (126);
		\draw (126) to (125);
		\draw (135) to (112);
		\draw (137) to (146.center);
		\draw (147.center) to (114);
		\draw (148) to (149);
		\draw (149) to (12);
		\draw (41) to (150);
		\draw (150) to (151);
		\draw (156) to (158);
		\draw (81) to (159);
	\end{pgfonlayer}
\end{tikzpicture}

One can use the bigraph algorithm to form quantum circuits of the semantic side of the meaning. In the pregroup type of the sentence ‘Sara ketab ra kharid’ set $o=n$. For atomic types $n$ and $s$ consider two qubits and one qubit respectively. The number of qubits for each type $t$ is the sum of the number of qubits associated to all atomic types in $t$. 
For example the transitive verb ‘kharid’ has five qubits.
For each word in the sentence we have a quantum circuit as follows:

\includegraphics[scale=0.5]{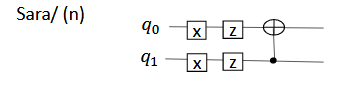}

\vspace{1cm}
\includegraphics[scale=0.5]{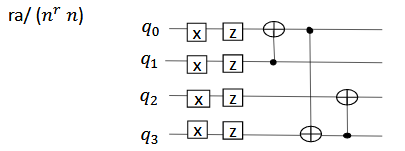}

\vspace{1cm}
\includegraphics[scale=0.5]{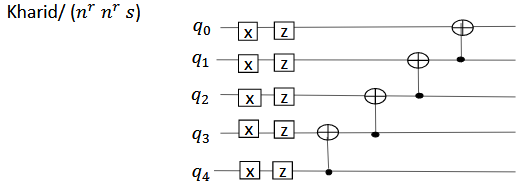}

The quantum circuit of the whole sentence is as follows:

\vspace{.5cm}
\includegraphics[scale=0.5]{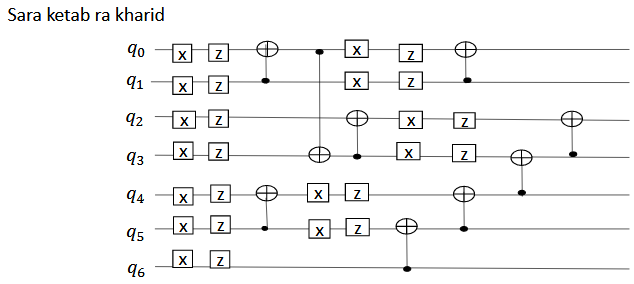}

The reduction diagram of the sentence ‘sara bought the book’ in English is:

\begin{tikzpicture}
	\begin{pgfonlayer}{nodelayer}
		\node [style=none] (0) at (-4.5, 4) {};
		\node [style=none] (1) at (-3, 4) {};
		\node [style=none] (2) at (-2.5, 4) {};
		\node [style=none] (3) at (-2, 4) {};
		\node [style=none] (4) at (-0.5, 4) {};
		\node [style=none] (5) at (0, 4) {};
		\node [style=none] (6) at (1.5, 4) {};
		\node [style=none] (7) at (-2.5, 3.25) {};
		\node [style=none] (8) at (-4.5, 4.25) {$n$};
		\node [style=none] (9) at (-3, 4.25) {$n^r$};
		\node [style=none] (10) at (-2.5, 4.25) {$s$};
		\node [style=none] (11) at (-2, 4.25) {$n^l$};
		\node [style=none] (12) at (-0.5, 4.25) {$n$};
		\node [style=none] (13) at (0, 4.25) {$n^l$};
		\node [style=none] (14) at (1.5, 4.25) {$n$};
	\end{pgfonlayer}
	\begin{pgfonlayer}{edgelayer}
		\draw [in=-90, out=-90, looseness=1.25] (0.center) to (1.center);
		\draw [in=-90, out=-90, looseness=1.25] (3.center) to (4.center);
		\draw [in=-90, out=-90, looseness=1.25] (5.center) to (6.center);
		\draw (2.center) to (7.center);
	\end{pgfonlayer}
\end{tikzpicture}

So the quantum circuit of the whole sentence is as follows:

\vspace{.5cm}
\includegraphics[scale=0.5]{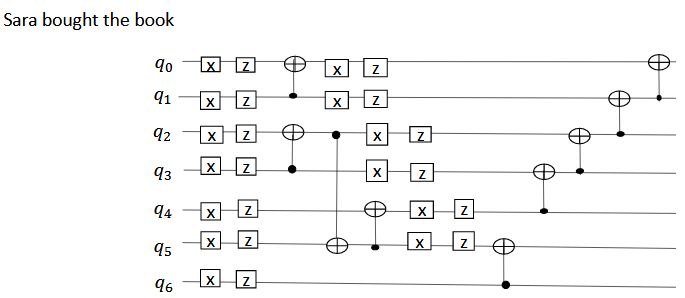}

The two sentences ‘Sara Ketab ra kharid’ and ‘Sara bought the book’ have the same meaning but are grammatically different. We expect the above two circuits to have the same output. 
According to \cite{f} we present grammar+meaning as quantum circuit for the above two sentences. Consider the states $\ket{\psi_{n_s}}$ and $\ket{\psi_{n_o}}$ correspond to the subject and the object, respectively. Also a transitive verb as a map $\eta_{tv}$ that takes $\ket{\psi_{n_s}} \in \mathbb{C}^2$ and $\ket{\psi_{n_o}} \in \mathbb{C}^2$ and produces $\ket{\psi_{n_s.n_o.tv}}\in \mathbb{C}^{2k}$, diagrammatically:

\begin{tikzpicture}
	\begin{pgfonlayer}{nodelayer}
		\node [style=none] (1) at (-5.75, 0.75) {};
		\node [style=none] (2) at (-3.75, 0.75) {};
		\node [style=none] (3) at (-4.75, 1.5) {};
		\node [style=none] (4) at (-4.75, 1) {$kharid$};
		\node [style=none] (5) at (-5.75, 1.25) {};
		\node [style=none] (6) at (-3.75, 1.25) {};
		\node [style=none] (7) at (-5.5, 0) {};
		\node [style=none] (9) at (-4, 0) {};
		\node [style=none] (10) at (-5.5, 0.75) {};
		\node [style=none] (12) at (-4, 0.75) {};
		\node [style=none] (13) at (-4.75, 0.75) {};
		\node [style=none] (14) at (-4.75, 0) {};
		\node [style=none] (16) at (-3, 1) {$=$};
		\node [style=none] (18) at (-1, 0.75) {};
		\node [style=none] (19) at (1, 1.25) {};
		\node [style=none] (20) at (1, 0.75) {};
		\node [style=none] (22) at (-1, 1.25) {};
		\node [style=none] (23) at (-0.5, 1.25) {};
		\node [style=none] (24) at (0, 1.25) {};
		\node [style=none] (25) at (0.5, 0.75) {};
		\node [style=none] (26) at (0.5, 0) {};
		\node [style=none] (27) at (-1.75, 0) {};
		\node [style=none] (28) at (-1.75, 0) {};
		\node [style=none] (29) at (-2.25, 0) {};
		\node [style=none] (30) at (0, 1) {};
		\node [style=none] (31) at (0, 1) {};
		\node [style=none] (32) at (0, 1) {};
		\node [style=none] (33) at (0, 1) {};
		\node [style=none] (34) at (0, 1) {$kharid$};
	\end{pgfonlayer}
	\begin{pgfonlayer}{edgelayer}
		\draw (5.center) to (1.center);
		\draw (6.center) to (2.center);
		\draw (5.center) to (3.center);
		\draw (3.center) to (6.center);
		\draw (10.center) to (7.center);
		\draw[line width=2pt] (12.center) to (9.center);
		\draw (1.center) to (2.center);
		\draw (13.center) to (14.center);
		\draw (18.center) to (20.center);
		\draw (19.center) to (20.center);
		\draw (22.center) to (19.center);
		\draw (22.center) to (18.center);
		\draw [in=90, out=120, looseness=1.75] (23.center) to (28.center);
		\draw [in=90, out=120, looseness=1.75] (24.center) to (29.center);
		\draw[line width=2pt] (25.center) to (26.center);
	\end{pgfonlayer}
\end{tikzpicture}

So $\ket{\psi_{kharid}} \in \mathbb{C}^{2}  \otimes \mathbb{C}^{2} \otimes \mathbb{C}^{2k}$.
Because the quantum model relies on the tensor product, an exponential blow-up occurs for meaning spaces of words. In order to avoid this obstacle in experiments decrease the dimension of the spaces in which meanings of transitive verbs live. 
For the transitive verb, instead of state in the large space $\ket{\psi_{kharid}} \in \mathbb{C}^{2}  \otimes \mathbb{C}^{2} \otimes \mathbb{C}^{2k}$
consider state in a smaller space
$\ket{\psi_{*kharid*}} \in \mathbb{C}^{2}  \otimes \mathbb{C}^{2}$, diagrammatically:

\begin{tikzpicture}
	\begin{pgfonlayer}{nodelayer}
		\node [style=none] (12) at (-2.5, 1.5) {};
		\node [style=none] (13) at (-2.5, 1.5) {$\mapsto$};
		\node [style=none] (16) at (-5.25, 1.25) {};
		\node [style=none] (17) at (-3.25, 1.25) {};
		\node [style=none] (18) at (-4.25, 2) {};
		\node [style=none] (19) at (-4.25, 1.5) {$kharid$};
		\node [style=none] (20) at (-5.25, 1.75) {};
		\node [style=none] (21) at (-3.25, 1.75) {};
		\node [style=none] (22) at (-5, 0.5) {};
		\node [style=none] (23) at (-3.5, 0.5) {};
		\node [style=none] (24) at (-5, 1.25) {};
		\node [style=none] (25) at (-3.5, 1.25) {};
		\node [style=none] (26) at (-4.25, 1.25) {};
		\node [style=none] (27) at (-4.25, 0.5) {};
		\node [style=none] (28) at (-1.75, 1.25) {};
		\node [style=none] (29) at (0.25, 1.25) {};
		\node [style=none] (30) at (-0.75, 2) {};
		\node [style=none] (31) at (-0.75, 1.5) {$*kharid*$};
		\node [style=none] (32) at (-1.75, 1.75) {};
		\node [style=none] (33) at (0.25, 1.75) {};
		\node [style=none] (34) at (-1.25, 0.5) {};
		\node [style=none] (35) at (-0.5, 0.5) {};
		\node [style=none] (36) at (-1.25, 1.25) {};
		\node [style=none] (37) at (-0.5, 1.25) {};
	\end{pgfonlayer}
	\begin{pgfonlayer}{edgelayer}
		\draw (20.center) to (16.center);
		\draw (21.center) to (17.center);
		\draw (20.center) to (18.center);
		\draw (18.center) to (21.center);
		\draw (24.center) to (22.center);
		\draw[line width=2pt] (25.center) to (23.center);
		\draw (16.center) to (17.center);
		\draw (26.center) to (27.center);
		\draw (32.center) to (28.center);
		\draw (33.center) to (29.center);
		\draw (32.center) to (30.center);
		\draw (30.center) to (33.center);
		\draw (36.center) to (34.center);
		\draw (37.center) to (35.center);
		\draw (28.center) to (29.center);
	\end{pgfonlayer}
\end{tikzpicture}

Then copy each of the wires and bundle two of the wires together to make up the thick wire. Thus ‘kharid’ is obtained:

\begin{tikzpicture}
	\begin{pgfonlayer}{nodelayer}
		\node [style=none] (0) at (-5, 3) {};
		\node [style=none] (1) at (-5, 2.5) {};
		\node [style=none] (2) at (-2.5, 3) {};
		\node [style=none] (3) at (-2.5, 2.5) {};
		\node [style=none] (4) at (-3.75, 3.25) {};
		\node [style=none] (5) at (-3.75, 2.75) {$*kharid*$};
		\node [style=none] (6) at (-4.25, 2.5) {};
		\node [style=none] (7) at (-3.25, 2.5) {};
		\node [style=new style 1] (8) at (-4.25, 1.75) {};
		\node [style=new style 1] (9) at (-3.25, 1.75) {};
		\node [style=none] (10) at (-5, 1) {};
		\node [style=none] (12) at (-2.5, 1) {};
		\node [style=none] (13) at (-4, 1) {};
		\node [style=none] (14) at (-3.75, 0.25) {};
		\node [style=none] (15) at (-2.25, 0.25) {};
		\node [style=none] (16) at (-3, -0.25) {};
		\node [style=none] (17) at (-3, -0.5) {};
		\node [style=none] (18) at (-3, -1.25) {};
		\node [style=none] (19) at (-5, 0.25) {};
		\node [style=none] (20) at (-4, 0.25) {};
		\node [style=none] (21) at (-2.5, 0.25) {};
		\node [style=none] (22) at (-3.5, 1) {};
		\node [style=none] (23) at (-3.5, 0.25) {};
	\end{pgfonlayer}
	\begin{pgfonlayer}{edgelayer}
		\draw (0.center) to (1.center);
		\draw (1.center) to (3.center);
		\draw (3.center) to (2.center);
		\draw (4.center) to (0.center);
		\draw (4.center) to (2.center);
		\draw (6.center) to (8);
		\draw (7.center) to (9);
		\draw [in=90, out=-180] (8) to (10.center);
		\draw [bend right=45] (9) to (13.center);
		\draw [bend left=45] (9) to (12.center);
		\draw [in=135, out=-90, looseness=1.25] (14.center) to (16.center);
		\draw [in=60, out=-105] (15.center) to (16.center);
		\draw[line width=2pt] (17.center) to (18.center);
		\draw (10.center) to (19.center);
		\draw (13.center) to (20.center);
		\draw (12.center) to (21.center);
		\draw [in=90, out=0] (8) to (22.center);
		\draw (22.center) to (23.center);
	\end{pgfonlayer}
\end{tikzpicture}
	
For more details see \cite{f}. Now inter Sara and Ketab into the picture:

\vspace{1cm}
\begin{tikzpicture}
	\begin{pgfonlayer}{nodelayer}
		\node [style=none] (0) at (-2.75, 3) {};
		\node [style=none] (1) at (-2.75, 2.5) {};
		\node [style=none] (2) at (-0.75, 3) {};
		\node [style=none] (3) at (-0.75, 2.5) {};
		\node [style=none] (4) at (-1.75, 3.25) {};
		\node [style=none] (5) at (-1.75, 2.75) {$*kharid*$};
		\node [style=none] (6) at (-2.25, 2.5) {};
		\node [style=none] (7) at (-1.25, 2.5) {};
		\node [style=new style 1] (8) at (-2.25, 2) {};
		\node [style=new style 1] (9) at (-1.25, 2) {};
		\node [style=none] (11) at (-1.75, 0.25) {};
		\node [style=none] (12) at (-0.75, 0.25) {};
		\node [style=none] (16) at (-5.25, 1.5) {};
		\node [style=none] (17) at (-5.25, 1) {};
		\node [style=none] (18) at (-4.25, 1.5) {};
		\node [style=none] (19) at (-4.25, 1) {};
		\node [style=none] (20) at (-4.75, 1.75) {};
		\node [style=none] (21) at (-4.75, 1.25) {$Ketab$};
		\node [style=none] (24) at (-7, 1.5) {};
		\node [style=none] (25) at (-7, 1) {};
		\node [style=none] (26) at (-6, 1.5) {};
		\node [style=none] (27) at (-6, 1) {};
		\node [style=none] (28) at (-6.5, 1.75) {};
		\node [style=none] (29) at (-6.5, 1.25) {$Sara$};
		\node [style=none] (30) at (-6.5, 1) {};
		\node [style=none] (31) at (-4.75, 1) {};
		\node [style=none] (33) at (0, 2) {$=$};
		\node [style=none] (34) at (5, 3) {};
		\node [style=none] (35) at (5, 2.5) {};
		\node [style=none] (36) at (6.5, 3) {};
		\node [style=none] (37) at (6.5, 2.5) {};
		\node [style=none] (38) at (5.75, 3.25) {};
		\node [style=none] (39) at (5.75, 2.75) {$*kharid*$};
		\node [style=none] (40) at (5.5, 2.5) {};
		\node [style=none] (41) at (6, 2.5) {};
		\node [style=none] (42) at (3, 3) {};
		\node [style=none] (43) at (3, 2.5) {};
		\node [style=none] (44) at (4, 3) {};
		\node [style=none] (45) at (4, 2.5) {};
		\node [style=none] (46) at (3.5, 3.25) {};
		\node [style=none] (47) at (3.5, 2.75) {$Ketab$};
		\node [style=none] (48) at (3.5, 2.5) {};
		\node [style=none] (49) at (0.75, 3) {};
		\node [style=none] (50) at (0.75, 2.5) {};
		\node [style=none] (51) at (1.75, 3) {};
		\node [style=none] (52) at (1.75, 2.5) {};
		\node [style=none] (53) at (1.25, 3.25) {};
		\node [style=none] (54) at (1.25, 2.75) {$Sara$};
		\node [style=none] (55) at (1.25, 2.5) {};
		\node [style=new style 1] (56) at (1.25, 0.5) {};
		\node [style=new style 1] (57) at (3.5, 1.5) {};
		\node [style=none] (58) at (1.25, -0.5) {};
		\node [style=none] (59) at (3.5, -0.5) {};
		\node [style=none] (62) at (-2, 1) {};
		\node [style=none] (66) at (-4.75, 1) {};
		\node [style=none] (67) at (-4.75, 1) {};
		\node [style=none] (68) at (-3, 1) {};
	\end{pgfonlayer}
	\begin{pgfonlayer}{edgelayer}
		\draw (0.center) to (1.center);
		\draw (1.center) to (3.center);
		\draw (3.center) to (2.center);
		\draw (4.center) to (0.center);
		\draw (4.center) to (2.center);
		\draw (6.center) to (8);
		\draw (7.center) to (9);
		\draw [in=90, out=-30, looseness=0.75] (8) to (11.center);
		\draw [in=90, out=-30] (9) to (12.center);
		\draw (16.center) to (17.center);
		\draw (17.center) to (19.center);
		\draw (19.center) to (18.center);
		\draw (20.center) to (16.center);
		\draw (20.center) to (18.center);
		\draw (24.center) to (25.center);
		\draw (25.center) to (27.center);
		\draw (27.center) to (26.center);
		\draw (28.center) to (24.center);
		\draw (28.center) to (26.center);
		\draw (34.center) to (35.center);
		\draw (35.center) to (37.center);
		\draw (37.center) to (36.center);
		\draw (38.center) to (34.center);
		\draw (38.center) to (36.center);
		\draw (42.center) to (43.center);
		\draw (43.center) to (45.center);
		\draw (45.center) to (44.center);
		\draw (46.center) to (42.center);
		\draw (46.center) to (44.center);
		\draw (49.center) to (50.center);
		\draw (50.center) to (52.center);
		\draw (52.center) to (51.center);
		\draw (53.center) to (49.center);
		\draw (53.center) to (51.center);
		\draw [in=0, out=-120] (40.center) to (57);
		\draw [in=-15, out=-105, looseness=0.75] (41.center) to (56);
		\draw (55.center) to (56);
		\draw (48.center) to (57);
		\draw (56) to (58.center);
		\draw (57) to (59.center);
		\draw [in=75, out=-165] (9) to (62.center);
		\draw (31.center) to (67.center);
		\draw [in=75, out=-165] (8) to (68.center);
		\draw [in=-90, out=-90, looseness=1.50] (67.center) to (68.center);
		\draw [in=-90, out=-90] (30.center) to (62.center);
	\end{pgfonlayer}
\end{tikzpicture}

We Pull some spiders out:

\begin{tikzpicture}
	\begin{pgfonlayer}{nodelayer}
		\node [style=none] (0) at (-0.5, 4) {};
		\node [style=none] (1) at (-0.5, 3.5) {};
		\node [style=none] (2) at (1, 4) {};
		\node [style=none] (3) at (1, 3.5) {};
		\node [style=none] (4) at (0.25, 4.25) {};
		\node [style=none] (5) at (0.25, 3.75) {$*kharid*$};
		\node [style=none] (6) at (-0.25, 3.5) {};
		\node [style=none] (7) at (0.75, 3.5) {};
		\node [style=none] (8) at (-3, 4) {};
		\node [style=none] (9) at (-3, 3.5) {};
		\node [style=none] (10) at (-2, 4) {};
		\node [style=none] (11) at (-2, 3.5) {};
		\node [style=none] (12) at (-2.5, 4.25) {};
		\node [style=none] (13) at (-2.5, 3.75) {$Ketab$};
		\node [style=none] (14) at (-4.5, 4) {};
		\node [style=none] (15) at (-4.5, 3.5) {};
		\node [style=none] (16) at (-3.5, 4) {};
		\node [style=none] (17) at (-3.5, 3.5) {};
		\node [style=none] (18) at (-4, 4.25) {};
		\node [style=none] (19) at (-4, 3.75) {$Sara$};
		\node [style=none] (20) at (-4, 3.5) {};
		\node [style=none] (21) at (-2.5, 3.5) {};
		\node [style=new style 1] (26) at (-2.5, 2.5) {};
		\node [style=new style 1] (27) at (-4, 1.5) {};
		\node [style=none] (28) at (-4, 0.75) {};
		\node [style=none] (29) at (-2.5, 0.75) {};
		\node [style=black dot] (30) at (-0.25, 2.5) {};
		\node [style=black dot] (31) at (-0.25, 0.75) {};
		\node [style=black dot] (33) at (0.75, 0.75) {};
		\node [style=black dot] (34) at (0.75, 1.5) {};
		\node [style=none] (35) at (2.25, 3.25) {$(4)$};
	\end{pgfonlayer}
	\begin{pgfonlayer}{edgelayer}
		\draw (0.center) to (1.center);
		\draw (1.center) to (3.center);
		\draw (3.center) to (2.center);
		\draw (4.center) to (0.center);
		\draw (4.center) to (2.center);
		\draw (8.center) to (9.center);
		\draw (9.center) to (11.center);
		\draw (11.center) to (10.center);
		\draw (12.center) to (8.center);
		\draw (12.center) to (10.center);
		\draw (14.center) to (15.center);
		\draw (15.center) to (17.center);
		\draw (17.center) to (16.center);
		\draw (18.center) to (14.center);
		\draw (18.center) to (16.center);
		\draw (20.center) to (27);
		\draw (21.center) to (26);
		\draw (26) to (29.center);
		\draw (27) to (28.center);
		\draw (6.center) to (30);
		\draw (30) to (31);
		\draw (30) to (26);
		\draw (27) to (34);
		\draw (7.center) to (34);
		\draw (34) to (33);
	\end{pgfonlayer}
\end{tikzpicture}
	
and by Using the Choi-Jamiolkowski correspondence we obtain:

\begin{tikzpicture}
	\begin{pgfonlayer}{nodelayer}
		\node [style=none] (0) at (4, 2) {};
		\node [style=none] (1) at (4, 1.5) {};
		\node [style=none] (2) at (5.5, 2) {};
		\node [style=none] (3) at (5.5, 1.5) {};
		\node [style=none] (4) at (4.75, 1.75) {$*kharid*$};
		\node [style=none] (5) at (4.25, 1.5) {};
		\node [style=none] (6) at (5.25, 1.5) {};
		\node [style=none] (7) at (1.5, 4) {};
		\node [style=none] (8) at (1.5, 3.5) {};
		\node [style=none] (9) at (2.5, 4) {};
		\node [style=none] (10) at (2.5, 3.5) {};
		\node [style=none] (11) at (2, 4.25) {};
		\node [style=none] (12) at (2, 3.75) {$Ketab$};
		\node [style=none] (13) at (0, 4) {};
		\node [style=none] (14) at (0, 3.5) {};
		\node [style=none] (15) at (1, 4) {};
		\node [style=none] (16) at (1, 3.5) {};
		\node [style=none] (17) at (0.5, 4.25) {};
		\node [style=none] (18) at (0.5, 3.75) {$Sara$};
		\node [style=none] (19) at (0.5, 3.5) {};
		\node [style=none] (20) at (2, 3.5) {};
		\node [style=new style 1] (24) at (2, 2.75) {};
		\node [style=new style 1] (25) at (0.5, 0.25) {};
		\node [style=none] (26) at (4.75, 1.5) {};
		\node [style=none] (27) at (0.5, -0.75) {};
		\node [style=none] (28) at (2, -0.75) {};
		\node [style=none] (30) at (4.75, 2) {};
		\node [style=black dot] (31) at (4.75, 2.75) {};
		\node [style=black dot] (32) at (4.75, 3.75) {};
		\node [style=black dot] (33) at (4.75, 0.25) {};
		\node [style=black dot] (34) at (4.75, -0.75) {};
		\node [style=none] (35) at (-2.25, 3) {};
		\node [style=none] (36) at (-2.25, 2.5) {};
		\node [style=none] (37) at (-0.75, 3) {};
		\node [style=none] (38) at (-0.75, 2.5) {};
		\node [style=none] (39) at (-1.5, 2.75) {$*kharid*$};
		\node [style=none] (40) at (-2, 2.5) {};
		\node [style=none] (41) at (-1, 2.5) {};
		\node [style=none] (42) at (-4.75, 3) {};
		\node [style=none] (43) at (-4.75, 2.5) {};
		\node [style=none] (44) at (-3.75, 3) {};
		\node [style=none] (45) at (-3.75, 2.5) {};
		\node [style=none] (46) at (-4.25, 3.25) {};
		\node [style=none] (47) at (-4.25, 2.75) {$Ketab$};
		\node [style=none] (48) at (-6.25, 3) {};
		\node [style=none] (49) at (-6.25, 2.5) {};
		\node [style=none] (50) at (-5.25, 3) {};
		\node [style=none] (51) at (-5.25, 2.5) {};
		\node [style=none] (52) at (-5.75, 3.25) {};
		\node [style=none] (53) at (-5.75, 2.75) {$Sara$};
		\node [style=none] (54) at (-5.75, 2.5) {};
		\node [style=none] (55) at (-4.25, 2.5) {};
		\node [style=new style 1] (56) at (-4.25, 1.25) {};
		\node [style=new style 1] (57) at (-5.75, 0.5) {};
		\node [style=none] (58) at (-1.5, 2.5) {};
		\node [style=none] (59) at (-5.75, -0.5) {};
		\node [style=none] (60) at (-4.25, -0.5) {};
		\node [style=none] (61) at (-1.5, 3) {};
		\node [style=none] (62) at (-2, 4.25) {};
		\node [style=none] (63) at (-3, 2.75) {};
		\node [style=black dot] (64) at (-3, 1.25) {};
		\node [style=black dot] (65) at (-3, -0.5) {};
		\node [style=black dot] (66) at (-1.5, -0.5) {};
		\node [style=black dot] (67) at (-1.5, 0.5) {};
		\node [style=none] (68) at (-0.25, 2.25) {};
		\node [style=none] (69) at (-0.25, 2.25) {$=$};
		\node [style=none] (70) at (6, 3.5) {$(5)$};
	\end{pgfonlayer}
	\begin{pgfonlayer}{edgelayer}
		\draw (0.center) to (1.center);
		\draw (1.center) to (3.center);
		\draw (3.center) to (2.center);
		\draw (7.center) to (8.center);
		\draw (8.center) to (10.center);
		\draw (10.center) to (9.center);
		\draw (11.center) to (7.center);
		\draw (11.center) to (9.center);
		\draw (13.center) to (14.center);
		\draw (14.center) to (16.center);
		\draw (16.center) to (15.center);
		\draw (17.center) to (13.center);
		\draw (17.center) to (15.center);
		\draw (19.center) to (25);
		\draw (20.center) to (24);
		\draw (25) to (27.center);
		\draw (24) to (28.center);
		\draw (0.center) to (2.center);
		\draw (24) to (31);
		\draw (32) to (31);
		\draw (31) to (30.center);
		\draw (26.center) to (33);
		\draw (25) to (33);
		\draw (33) to (34);
		\draw (35.center) to (36.center);
		\draw (36.center) to (38.center);
		\draw (38.center) to (37.center);
		\draw (42.center) to (43.center);
		\draw (43.center) to (45.center);
		\draw (45.center) to (44.center);
		\draw (46.center) to (42.center);
		\draw (46.center) to (44.center);
		\draw (48.center) to (49.center);
		\draw (49.center) to (51.center);
		\draw (51.center) to (50.center);
		\draw (52.center) to (48.center);
		\draw (52.center) to (50.center);
		\draw (54.center) to (57);
		\draw (55.center) to (56);
		\draw (57) to (59.center);
		\draw (56) to (60.center);
		\draw (35.center) to (37.center);
		\draw [in=-30, out=90] (61.center) to (62.center);
		\draw [in=90, out=165, looseness=1.25] (62.center) to (63.center);
		\draw (63.center) to (64);
		\draw (64) to (65);
		\draw (58.center) to (67);
		\draw (67) to (66);
		\draw (67) to (57);
		\draw (64) to (56);
	\end{pgfonlayer}
\end{tikzpicture}

The circuit (4) requires 4 qubits and has two CNOT-gates in parallel, but the circuit (5) requires 3 qubits and has sequential CNOT-gates.
Indeed, the use of the Choi-Jamiolkowski correspondence has reduced the number of qubits, but has increased the depth of the CNOT-gates.
As mentioned in \cite{f} ion trap hardware has less qubits, but performs better for greater circuit depth.
In ZX-calculus and via Euler decomposition any one-qubit unitary gate is represented as follows:

\begin{center}
\begin{tikzpicture}
	\begin{pgfonlayer}{nodelayer}
		\node [style=new style 2] (0) at (0.5, 1.75) {};
		\node [style=new style 2] (1) at (0.5, 1.75) {$\beta$};
		\node [style=new style 3] (4) at (0.5, 2.75) {$\alpha$};
		\node [style=new style 3] (5) at (0.5, 0.75) {$\gamma$};
		\node [style=none] (6) at (0.5, 3.5) {};
		\node [style=none] (7) at (0.5, 0) {};
	\end{pgfonlayer}
	\begin{pgfonlayer}{edgelayer}
		\draw (4) to (1);
		\draw (1) to (5);
		\draw (6.center) to (4);
		\draw (5) to (7.center);
	\end{pgfonlayer}
\end{tikzpicture}
\end{center}
Each verb is represented by an unitary gate $U$ and has different values $\alpha, \beta$ and $\gamma$.
So we obtain:

\begin{tikzpicture}
	\begin{pgfonlayer}{nodelayer}
		\node [style=none] (0) at (-1.5, 2) {};
		\node [style=none] (1) at (-1.5, 1.5) {};
		\node [style=none] (2) at (-1, 2) {};
		\node [style=none] (3) at (-1, 1.5) {};
		\node [style=none] (4) at (-1.25, 1.75) {$U$};
		\node [style=none] (5) at (-1.5, 1.5) {};
		\node [style=none] (6) at (-1, 1.5) {};
		\node [style=none] (7) at (-3.75, 4.75) {};
		\node [style=none] (8) at (-3.75, 4.25) {};
		\node [style=none] (9) at (-2.75, 4.75) {};
		\node [style=none] (10) at (-2.75, 4.25) {};
		\node [style=none] (11) at (-3.25, 5) {};
		\node [style=none] (12) at (-3.25, 4.5) {$Ketab$};
		\node [style=none] (13) at (-5.75, 4.75) {};
		\node [style=none] (14) at (-5.75, 4.25) {};
		\node [style=none] (15) at (-4.75, 4.75) {};
		\node [style=none] (16) at (-4.75, 4.25) {};
		\node [style=none] (17) at (-5.25, 5) {};
		\node [style=none] (18) at (-5.25, 4.5) {$Sara$};
		\node [style=none] (19) at (-5.25, 4.25) {};
		\node [style=none] (20) at (-3.25, 4.25) {};
		\node [style=new style 1] (21) at (-3.25, 3.5) {};
		\node [style=new style 1] (22) at (-5.25, -0.25) {};
		\node [style=none] (23) at (-5.25, -1) {};
		\node [style=none] (24) at (-3.25, -1) {};
		\node [style=none] (25) at (-1.25, 2) {};
		\node [style=black dot] (26) at (-1.25, 4.5) {};
		\node [style=black dot] (27) at (-1.25, 3.5) {};
		\node [style=black dot] (28) at (-1.25, -0.25) {};
		\node [style=black dot] (29) at (-1.25, -1) {};
		\node [style=none] (30) at (-1.25, 1.5) {};
		\node [style=none] (31) at (2.25, 4.75) {};
		\node [style=none] (32) at (2.25, 4.25) {};
		\node [style=none] (33) at (3.25, 4.75) {};
		\node [style=none] (34) at (3.25, 4.25) {};
		\node [style=none] (35) at (2.75, 5) {};
		\node [style=none] (36) at (2.75, 4.5) {$Ketab$};
		\node [style=none] (37) at (0.25, 4.75) {};
		\node [style=none] (38) at (0.25, 4.25) {};
		\node [style=none] (39) at (1.25, 4.75) {};
		\node [style=none] (40) at (1.25, 4.25) {};
		\node [style=none] (41) at (0.75, 5) {};
		\node [style=none] (42) at (0.75, 4.5) {$Sara$};
		\node [style=none] (43) at (0.75, 4.25) {};
		\node [style=none] (44) at (2.75, 4.25) {};
		\node [style=new style 1] (45) at (2.75, 3.5) {};
		\node [style=new style 1] (46) at (0.75, -0.25) {};
		\node [style=none] (47) at (0.75, -1) {};
		\node [style=none] (48) at (2.75, -1) {};
		\node [style=new style 2] (49) at (5, 4.5) {};
		\node [style=black dot] (50) at (5, 4.5) {};
		\node [style=black dot] (51) at (5, 3.5) {};
		\node [style=black dot] (52) at (5, -0.25) {};
		\node [style=black dot] (53) at (5, -1) {};
		\node [style=none] (60) at (-0.25, 1.75) {};
		\node [style=none] (61) at (-0.25, 1.75) {$=$};
		\node [style=new style 2] (62) at (5, 1.75) {};
		\node [style=new style 2] (63) at (5, 1.75) {$\beta$};
		\node [style=new style 3] (64) at (5, 2.75) {$\alpha$};
		\node [style=new style 3] (65) at (5, 0.75) {$\gamma$};
	\end{pgfonlayer}
	\begin{pgfonlayer}{edgelayer}
		\draw (0.center) to (1.center);
		\draw (1.center) to (3.center);
		\draw (3.center) to (2.center);
		\draw (7.center) to (8.center);
		\draw (8.center) to (10.center);
		\draw (10.center) to (9.center);
		\draw (11.center) to (7.center);
		\draw (11.center) to (9.center);
		\draw (13.center) to (14.center);
		\draw (14.center) to (16.center);
		\draw (16.center) to (15.center);
		\draw (17.center) to (13.center);
		\draw (17.center) to (15.center);
		\draw (19.center) to (22);
		\draw (20.center) to (21);
		\draw (22) to (23.center);
		\draw (21) to (24.center);
		\draw (0.center) to (2.center);
		\draw (27) to (21);
		\draw (26) to (27);
		\draw (28) to (22);
		\draw (28) to (29);
		\draw (30.center) to (28);
		\draw (27) to (25.center);
		\draw (31.center) to (32.center);
		\draw (32.center) to (34.center);
		\draw (34.center) to (33.center);
		\draw (35.center) to (31.center);
		\draw (35.center) to (33.center);
		\draw (37.center) to (38.center);
		\draw (38.center) to (40.center);
		\draw (40.center) to (39.center);
		\draw (41.center) to (37.center);
		\draw (41.center) to (39.center);
		\draw (43.center) to (46);
		\draw (44.center) to (45);
		\draw (46) to (47.center);
		\draw (45) to (48.center);
		\draw (50) to (51);
		\draw (52) to (46);
		\draw (52) to (53);
		\draw (51) to (45);
		\draw (64) to (63);
		\draw (63) to (65);
		\draw (51) to (64);
		\draw (65) to (52);
	\end{pgfonlayer}
\end{tikzpicture}
 
By considering the singular value decomposition for the verb we obtain:

\begin{tikzpicture}
	\begin{pgfonlayer}{nodelayer}
		\node [style=none] (0) at (5.5, 3.5) {};
		\node [style=none] (1) at (5.5, 3) {};
		\node [style=none] (2) at (6.5, 3.5) {};
		\node [style=none] (3) at (6.5, 3) {};
		\node [style=none] (4) at (6, 3.75) {};
		\node [style=none] (5) at (6, 3.25) {$Ketab$};
		\node [style=none] (6) at (3.5, 3.5) {};
		\node [style=none] (7) at (3.5, 3) {};
		\node [style=none] (8) at (4.5, 3.5) {};
		\node [style=none] (9) at (4.5, 3) {};
		\node [style=none] (10) at (4, 3.75) {};
		\node [style=none] (11) at (4, 3.25) {$Sara$};
		\node [style=none] (12) at (4, 3) {};
		\node [style=none] (13) at (6, 3) {};
		\node [style=new style 1] (14) at (6, 1) {};
		\node [style=new style 1] (15) at (4, -6) {};
		\node [style=none] (16) at (4, -6.75) {};
		\node [style=new style 2] (18) at (7.75, 3.25) {};
		\node [style=black dot] (19) at (7.75, 1) {};
		\node [style=black dot] (20) at (7.75, -6) {};
		\node [style=black dot] (21) at (7.75, -6.75) {};
		\node [style=white dot] (24) at (7.75, -2.5) {};
		\node [style=black dot] (34) at (9.25, -2.5) {};
		\node [style=black dot] (35) at (9.25, -3.25) {};
		\node [style=none] (44) at (8.75, 3.5) {};
		\node [style=none] (45) at (8.75, 3) {};
		\node [style=none] (46) at (9.75, 3.5) {};
		\node [style=none] (47) at (9.75, 3) {};
		\node [style=none] (48) at (9.25, 3.75) {};
		\node [style=none] (49) at (9.25, 3.25) {$P$};
		\node [style=none] (50) at (9.25, 3) {};
		\node [style=new style 2] (56) at (7.75, -0.75) {};
		\node [style=new style 2] (57) at (7.75, -0.75) {$\beta$};
		\node [style=new style 3] (58) at (7.75, 0.25) {$\alpha$};
		\node [style=new style 3] (59) at (7.75, -1.75) {$\gamma$};
		\node [style=new style 2] (60) at (7.75, -4.25) {};
		\node [style=new style 2] (61) at (7.75, -4.25) {$\beta'$};
		\node [style=new style 3] (62) at (7.75, -3.25) {$\alpha'$};
		\node [style=new style 3] (63) at (7.75, -5.25) {$\gamma'$};
		\node [style=none] (64) at (6, -6.75) {};
	\end{pgfonlayer}
	\begin{pgfonlayer}{edgelayer}
		\draw (0.center) to (1.center);
		\draw (1.center) to (3.center);
		\draw (3.center) to (2.center);
		\draw (4.center) to (0.center);
		\draw (4.center) to (2.center);
		\draw (6.center) to (7.center);
		\draw (7.center) to (9.center);
		\draw (9.center) to (8.center);
		\draw (10.center) to (6.center);
		\draw (10.center) to (8.center);
		\draw (15) to (16.center);
		\draw (20) to (15);
		\draw (20) to (21);
		\draw (19) to (14);
		\draw (18) to (19);
		\draw (34) to (35);
		\draw (24) to (34);
		\draw (12.center) to (15);
		\draw (13.center) to (14);
		\draw (44.center) to (45.center);
		\draw (45.center) to (47.center);
		\draw (47.center) to (46.center);
		\draw (48.center) to (44.center);
		\draw (48.center) to (46.center);
		\draw (50.center) to (34);
		\draw (58) to (57);
		\draw (57) to (59);
		\draw (19) to (58);
		\draw (59) to (24);
		\draw (62) to (61);
		\draw (61) to (63);
		\draw (24) to (62);
		\draw (63) to (20);
		\draw (14) to (64.center);
	\end{pgfonlayer}
\end{tikzpicture}
		 
Where the state $P$ is the diagonal of the matrix.
We represent all noun states by gates:

\begin{tikzpicture}
	\begin{pgfonlayer}{nodelayer}
		\node [style=new style 1] (0) at (-1, 1) {};
		\node [style=new style 1] (1) at (-2.75, -4.75) {};
		\node [style=none] (2) at (-2.75, -5.75) {};
		\node [style=none] (3) at (-1, -5.75) {};
		\node [style=new style 2] (4) at (0.75, 3.75) {};
		\node [style=black dot] (5) at (0.75, 1) {};
		\node [style=black dot] (6) at (0.75, -4.75) {};
		\node [style=black dot] (7) at (0.75, -5.5) {};
		\node [style=black dot] (17) at (-2.75, 3.75) {};
		\node [style=black dot] (20) at (-1, 3.75) {};
		\node [style=white dot] (21) at (0.75, -1.25) {};
		\node [style=none] (22) at (0.25, -1.75) {};
		\node [style=none] (23) at (0.25, -2.25) {};
		\node [style=none] (24) at (1.25, -1.75) {};
		\node [style=none] (25) at (1.25, -2.25) {};
		\node [style=none] (33) at (0.75, -1.75) {};
		\node [style=none] (34) at (0.75, -2.25) {};
		\node [style=none] (35) at (0.75, -2) {};
		\node [style=none] (36) at (0.75, -2) {$\alpha'+\gamma$};
		\node [style=black dot] (37) at (2.25, 3.75) {};
		\node [style=black dot] (38) at (2.25, -1.25) {};
		\node [style=black dot] (39) at (2.25, -2) {};
		\node [style=black dot] (70) at (-2.75, 2.75) {};
		\node [style=black dot] (71) at (-2.75, 2.75) {$\alpha_s$};
		\node [style=white dot] (72) at (-2.75, 1.75) {};
		\node [style=white dot] (73) at (-2.75, 1.75) {$\beta_s$};
		\node [style=black dot] (74) at (-1, 2.75) {$\alpha_k$};
		\node [style=white dot] (76) at (-1, 1.75) {};
		\node [style=white dot] (77) at (-1, 1.75) {$\beta_k$};
		\node [style=white dot] (78) at (0.75, 0.25) {};
		\node [style=white dot] (79) at (0.75, 0.25) {$\alpha$};
		\node [style=black dot] (80) at (0.75, -0.5) {};
		\node [style=black dot] (81) at (0.75, -0.5) {$\beta$};
		\node [style=white dot] (84) at (0.75, -4) {};
		\node [style=white dot] (85) at (0.75, -4) {$\gamma'$};
		\node [style=black dot] (86) at (2.25, 2.75) {};
		\node [style=black dot] (87) at (2.25, 2.75) {$\alpha_p$};
		\node [style=black dot] (88) at (0.75, -3) {};
		\node [style=black dot] (89) at (0.75, -3) {$\beta'$};
	\end{pgfonlayer}
	\begin{pgfonlayer}{edgelayer}
		\draw (1) to (2.center);
		\draw (0) to (3.center);
		\draw (6) to (1);
		\draw (6) to (7);
		\draw (5) to (0);
		\draw (4) to (5);
		\draw [in=-180, out=15, looseness=0.75] (22.center) to (24.center);
		\draw [in=150, out=-150] (22.center) to (23.center);
		\draw [in=-165, out=-30, looseness=0.25] (23.center) to (25.center);
		\draw [in=30, out=-30, looseness=1.25] (24.center) to (25.center);
		\draw (21) to (33.center);
		\draw (38) to (39);
		\draw (21) to (38);
		\draw (17) to (71);
		\draw (71) to (73);
		\draw (73) to (1);
		\draw (20) to (74);
		\draw (74) to (77);
		\draw (77) to (0);
		\draw (5) to (79);
		\draw (79) to (81);
		\draw (81) to (21);
		\draw (85) to (6);
		\draw (37) to (87);
		\draw (87) to (38);
		\draw (34.center) to (89);
		\draw (89) to (85);
	\end{pgfonlayer}
\end{tikzpicture}	

For the sentence ‘sara bought the book’ we obtain the DisCoCat diagram as follows:

\begin{tikzpicture}
	\begin{pgfonlayer}{nodelayer}
		\node [style=none] (12) at (-2.75, 2.5) {};
		\node [style=none] (17) at (-4.5, 1.75) {};
		\node [style=none] (19) at (-6, 1.75) {};
		\node [style=none] (26) at (-2.75, 2) {$Bought$};
		\node [style=none] (33) at (1.75, 1.75) {};
		\node [style=none] (35) at (2.75, 2.5) {};
		\node [style=none] (37) at (2.75, 2) {$Book$};
		\node [style=none] (38) at (-4, 1.75) {};
		\node [style=none] (39) at (-1.5, 1.75) {};
		\node [style=none] (40) at (-3.25, 1.75) {};
		\node [style=none] (41) at (-5.25, 1.75) {};
		\node [style=none] (42) at (3.75, 1.75) {};
		\node [style=none] (43) at (-0.5, 1.75) {};
		\node [style=none] (44) at (1, 1.75) {};
		\node [style=none] (45) at (-2.25, 1.75) {};
		\node [style=none] (46) at (2.75, 1.75) {};
		\node [style=none] (47) at (-2.75, 1.75) {};
		\node [style=none] (48) at (-2.75, 1) {};
		\node [style=none] (49) at (-5.25, 2.5) {};
		\node [style=none] (50) at (-5.25, 2) {$Sara$};
	\end{pgfonlayer}
	\begin{pgfonlayer}{edgelayer}
		\draw (35.center) to (33.center);
		\draw (19.center) to (17.center);
		\draw (12.center) to (38.center);
		\draw (12.center) to (39.center);
		\draw (38.center) to (39.center);
		\draw [in=-90, out=-90, looseness=1.25] (41.center) to (40.center);
		\draw (35.center) to (42.center);
		\draw (33.center) to (42.center);
		\draw [in=-90, out=-90, looseness=1.50] (45.center) to (43.center);
		\draw [in=-90, out=-90, looseness=1.25] (46.center) to (44.center);
		\draw [in=90, out=90, looseness=1.50] (43.center) to (44.center);
		\draw (47.center) to (48.center);
		\draw (49.center) to (19.center);
		\draw (49.center) to (17.center);
	\end{pgfonlayer}
\end{tikzpicture}

that is equal to:

\begin{tikzpicture}
	\begin{pgfonlayer}{nodelayer}
		\node [style=none] (0) at (-4, 2.75) {};
		\node [style=none] (1) at (-5.5, 2) {};
		\node [style=none] (2) at (-7, 2) {};
		\node [style=none] (5) at (-4, 2.25) {$Bought$};
		\node [style=none] (6) at (-2.5, 2) {};
		\node [style=none] (7) at (-1.5, 2.75) {};
		\node [style=none] (9) at (-1.5, 2.25) {$Book$};
		\node [style=none] (10) at (-5.25, 2) {};
		\node [style=none] (11) at (-2.75, 2) {};
		\node [style=none] (12) at (-4.5, 2) {};
		\node [style=none] (13) at (-6.25, 2) {};
		\node [style=none] (14) at (-0.5, 2) {};
		\node [style=none] (16) at (-3.25, 2) {};
		\node [style=none] (18) at (-1.5, 2) {};
		\node [style=none] (19) at (-4, 2) {};
		\node [style=none] (20) at (-4, 1.25) {};
		\node [style=none] (21) at (-6.25, 2.75) {};
		\node [style=none] (22) at (-6.25, 2.25) {$Sara$};
		\node [style=none] (23) at (1, 2) {(6)};
	\end{pgfonlayer}
	\begin{pgfonlayer}{edgelayer}
		\draw (7.center) to (6.center);
		\draw (2.center) to (1.center);
		\draw (0.center) to (10.center);
		\draw (0.center) to (11.center);
		\draw (10.center) to (11.center);
		\draw [in=-90, out=-90, looseness=1.25] (13.center) to (12.center);
		\draw (7.center) to (14.center);
		\draw (6.center) to (14.center);
		\draw [in=-90, out=-90, looseness=1.25] (18.center) to (16.center);
		\draw (19.center) to (20.center);
		\draw (21.center) to (2.center);
		\draw (21.center) to (1.center);
	\end{pgfonlayer}
\end{tikzpicture}

Indeed we ignore ‘the’ and ‘ra’ of positive transitive sentences in English and Persian respectively. Therefore according to \cite{f} the parametrised quantum circuit of the diagram (6) is as follows:

\begin{tikzpicture}
 	\begin{pgfonlayer}{nodelayer}
 		\node [style=new style 1] (0) at (1.25, 2) {};
 		\node [style=new style 1] (1) at (-4, -3.75) {};
 		\node [style=none] (2) at (-4, -4.75) {};
 		\node [style=none] (3) at (1.25, -4.5) {};
 		\node [style=new style 2] (4) at (-0.5, 4.75) {};
 		\node [style=black dot] (5) at (-0.5, 2) {};
 		\node [style=black dot] (6) at (-0.5, -3.75) {};
 		\node [style=black dot] (7) at (-0.5, -4.5) {};
 		\node [style=black dot] (8) at (-4, 4.75) {};
 		\node [style=black dot] (9) at (1.25, 4.75) {};
 		\node [style=white dot] (10) at (-0.5, -0.25) {};
 		\node [style=none] (11) at (-1, -0.75) {};
 		\node [style=none] (12) at (-1, -1.25) {};
 		\node [style=none] (13) at (0, -0.75) {};
 		\node [style=none] (14) at (0, -1.25) {};
 		\node [style=none] (15) at (-0.5, -0.75) {};
 		\node [style=none] (16) at (-0.5, -1.25) {};
 		\node [style=none] (17) at (-0.5, -1) {};
 		\node [style=none] (18) at (-0.5, -1) {$\alpha'+\gamma$};
 		\node [style=black dot] (19) at (-2.25, 4.75) {};
 		\node [style=black dot] (20) at (-2.25, -0.25) {};
 		\node [style=black dot] (21) at (-2.25, -1) {};
 		\node [style=black dot] (22) at (-4, 3.75) {};
 		\node [style=black dot] (23) at (-4, 3.75) {$\alpha_S$};
 		\node [style=white dot] (24) at (-4, 2.75) {};
 		\node [style=white dot] (25) at (-4, 2.75) {$\beta_S$};
 		\node [style=black dot] (26) at (1.25, 3.75) {$\alpha_B$};
 		\node [style=white dot] (27) at (1.25, 2.75) {};
 		\node [style=white dot] (28) at (1.25, 2.75) {$\beta_B$};
 		\node [style=white dot] (29) at (-0.5, 1.25) {};
 		\node [style=white dot] (30) at (-0.5, 1.25) {$\alpha$};
 		\node [style=black dot] (31) at (-0.5, 0.5) {};
 		\node [style=black dot] (32) at (-0.5, 0.5) {$\beta$};
 		\node [style=white dot] (33) at (-0.5, -3) {};
 		\node [style=white dot] (34) at (-0.5, -3) {$\gamma'$};
 		\node [style=black dot] (35) at (-2.25, 3.75) {};
 		\node [style=black dot] (36) at (-2.25, 3.75) {$\alpha_P$};
 		\node [style=black dot] (37) at (-0.5, -2) {};
 		\node [style=black dot] (38) at (-0.5, -2) {$\beta'$};
 	\end{pgfonlayer}
 	\begin{pgfonlayer}{edgelayer}
 		\draw (1) to (2.center);
 		\draw (0) to (3.center);
 		\draw (6) to (1);
 		\draw (6) to (7);
 		\draw (5) to (0);
 		\draw (4) to (5);
 		\draw [in=-180, out=15, looseness=0.75] (11.center) to (13.center);
 		\draw [in=150, out=-150] (11.center) to (12.center);
 		\draw [in=-165, out=-30, looseness=0.25] (12.center) to (14.center);
 		\draw [in=30, out=-30, looseness=1.25] (13.center) to (14.center);
 		\draw (10) to (15.center);
 		\draw (20) to (21);
 		\draw (10) to (20);
 		\draw (8) to (23);
 		\draw (23) to (25);
 		\draw (25) to (1);
 		\draw (9) to (26);
 		\draw (26) to (28);
 		\draw (28) to (0);
 		\draw (5) to (30);
 		\draw (30) to (32);
 		\draw (32) to (10);
 		\draw (34) to (6);
 		\draw (19) to (36);
 		\draw (36) to (20);
 		\draw (16.center) to (38);
 		\draw (38) to (34);
 	\end{pgfonlayer}
 \end{tikzpicture}

\section{Conclusion and Future Works}
This paper extended the compact categorical semantics to analyse meanings of positive transitive sentences in Persian. It is necessary to introduce linear maps to represent the meaning of negative transitive sentences and grammatically more complex sentences in Persian. 
In this work the two sentences 'Sara ketab ra kharid' (in Persian) and 'Sara bought the book’ (in English) are instantiated as parametrised quantum circuits. The meaning of the two sentences are the same but the appearence of the obtained quantum circuits are different. These circuits need to be compiled correctly, thus it is necessary to introduce a test measurement at the terminal of the circuits to give almost similar results for the meaning of the synonymous sentences in different languages. As a future prospect one may use the compiler $t\ket{ket}$ to this aim, and run the circuits on the IBMQ and analyze the results. 


\begin{backmatter}


\section*{Authors' information}

mina.abbaszade@math.uk.ac.ir

vahid.salari@ehu.eus



\bibliographystyle{bmc-mathphys} 
\bibliography{bmc_article}      


\end{backmatter}
\end{document}